\documentclass[nohyper,12pt,letterpaper]{JHEP3}
\usepackage{epsfig,youngtab}

\newcommand{\myfig}[3]{\begin{figure}[ht]
\begin{center}
\leavevmode \epsfxsize=#2cm \epsfbox{#1}
\end{center}
\caption{#3} \label{fig:#1}
\end{figure}}

\author{Robert de Mello Koch$^{1,2}$ Tanay K. Dey$^{1}$, Norman Ives$^{1}$ and Michael Stephanou$^{1}$\\
\qquad \\
$^{1}$ National Institute for Theoretical Physics,\\
Department of Physics and Centre for Theoretical Physics,\\ 
University of the Witwatersrand,\\ 
Wits, 2050,\\ 
South Africa\\
\qquad\\
$^{2}$Stellenbosch Institute for Advanced Studies,\\
Stellenbosch,\\
South Africa\\
\qquad\\
E-mail: \email{robert@neo.phys.wits.ac.za, Tanay.Dey@wits.ac.za, Norman.Ives,Michael.Stephanou@students.wits.ac.za}}

\abstract{Correlation functions of operators with a conformal dimension of $O(N^2)$ are not well approximated by the planar limit.
The non-planar diagrams, which in the bulk spacetime correspond to string loop corrections, are enhanced by huge combinatorial factors. 
In this article we show how these loop corrections can be resummed. As a typical example of our results, in the half-BPS background of 
$M$ maximal giant gravitons we find the usual $1/N$ expansion is replaced by a $1/(M+N)$ expansion. Further, we find that there is a simple
exact relationship between amplitudes computed in the trivial background and amplitudes computed in the background of $M$ maximal
giant gravitons. Finally, we also find strong evidence for the BMN-type sectors suggested in arXiv:0801.4457.
The decoupling limit of arXiv:0801.4457 captures the decoupled low energy world volume theory of
the intersecting giant graviton system and this theory is weakly coupled even when the original ${\cal N}=4$ super Yang-Mills
theory is strongly coupled.}

\preprint{WITS-CTP-040}

\title{Correlators Of Operators with a Large {\cal R}-charge}


\def \Tr{\mbox{Tr\,}}

\begin{document}

\section{Introduction}

According to the AdS/CFT correspondence\cite{Maldacena:1997re}, the conformal dimension of an operator in the ${\cal N}=4$ super 
Yang-Mills theory maps into the energy of the corresponding state in IIB string theory on the AdS$_5\times$S$^5$ background. Thus, 
operators with a very large dimension will map into states with a very large energy. If this energy is large enough, back reaction 
can not be neglected and the state is best thought of as a new geometry which is only asymptotically AdS$_5\times$S$^5$. Good 
examples of such operators include the Schur polynomials\cite{Corley:2001zk,Corley:2002mj,Berenstein:2004kk} with ${\cal R}$-charge 
of $O(N^2)$ (dual to LLM geometries\cite{Lin:2004nb,Balasubramanian:2005mg}) and the operators obtained by distributing a gas of 
defects on Schur polynomials (which seem to be dual to asymptotically AdS$_5\times$S$^5$ charged black 
holes\cite{Balasubramanian:2007bs,Fareghbal:2008ar}). To build a proper understanding of these operators in the gauge theory one 
would like, at least, to compute the anomalous dimension of these operators and to deal with their mixing. 

In this article we consider the problem of computing correlators of heavy operators. The operators
we have in mind are a small perturbation of a BPS operator which has ${\cal R}$-charge $\sim$ conformal dimension
$\sim$ $N^2$. In the language of \cite{Berenstein:2003ah} we study {\it almost} BPS operators.
To solve this problem one must reorganize perturbation theory by resumming an infinite number of 
diagrams. The need for this reorganization is that for these operators non-planar diagrams can't be neglected\cite{Balasubramanian:2001nh}.
A similar situation is provided by the BMN sector\cite{Berenstein:2002jq} of ${\cal N}=4$ super Yang-Mills theory. In this case one considers operators 
with ${\cal R}$-charge $J\sim O(\sqrt{N})$. One finds the usual ${1\over N}$ expansion parameter is replaced by a new expansion 
parameter equal to ${J^2\over N}$\cite{Kristjansen:2002bb}. Thus, one must take ${J^2\over N}\ll 1$ to suppress non-planar diagrams.

The question we ask and answer in this paper is: 

{\vskip 0.5cm}

{\sl Is there a reorganization of the ${1\over N}$ expansion for almost BPS operators 
which have ${\cal R}$-charge of $O(N^2)$ and if so, what is the expansion parameter?}

{\vskip 0.5cm}

\noindent
We focus on almost BPS operators since we want to extrapolate our computations to strong coupling, where we can compare with the dual
string theory.

A particularly useful way to describe the half-BPS sector is to use the Schur polynomials\cite{Corley:2001zk,Corley:2002mj,Berenstein:2004kk}.
Among the operators we consider, are small perturbations of a Schur polynomial $\chi_B(Z)$ with $B$ a Young diagram that has $M$ columns
and $N$ rows, and $M$ is $O(N)$. This corresponds to an LLM geometry\cite{Lin:2004nb} with boundary condition that is an annulus.
The inner radius of the annulus is $\propto\sqrt{M}$ and the outer radius is $\propto\sqrt{M+N}$. We demonstrate, in section 2.1, that there is a 
reorganization of the perturbation theory in the half-BPS sector and that the new expansion parameter is ${1\over M+N}$. In section 2.2 we confirm
this answer by studying the holography of the IIB supergravity in the relevant LLM geometry. We then generalize these results to multi-ring
LLM geometries (in section 2.3), backgrounds with more than one charge (in sections 2.4 and 2.5) and beyond the BPS sector (in section 3). The 
Schwinger-Dyson equations provide a very powerful approach to the computation of correlators in the annulus background. We present these details
in the Appendices.

\section{Half-BPS Sector}

Schur polynomials provide a very convenient reorganization of the half-BPS sector. This is due to the fact that their two point
function is known to all orders in ${1\over N}$\cite{Corley:2001zk,Corley:2002mj} and that they satisfy a product rule allowing computation of
exact $n$-point correlators using only two-point functions (see Appendix D for a summary of the results we use). In this section 
we will use this Schur technology to provide an answer, in the half-BPS sector, to the question posed above.

${\cal N}=4$ super Yang-Mills theory has 6 scalars $\phi_i$ transforming in the adjoint of the gauge group and in the 
${\bf 6}$ of the $SU(4)_{\cal R}$ symmetry. We shall use the complex combinations
$$ Z=\phi_1+i\phi_2,\qquad Y=\phi_3+i\phi_4,\qquad X=\phi_5+i\phi_6,$$
in what follows.
We focus on the contributions coming from the color combinatorics; we drop all spacetime dependence from two point
correlators\footnote{It is simple to reinstate the spacetime dependence in the final result.}, that is, we use the two point functions
\begin{equation}
\left\langle Z_{ij}Z^\dagger_{kl}\right\rangle =\left\langle Y_{ij}Y^\dagger_{kl}\right\rangle =\left\langle X_{ij}X^\dagger_{kl}\right\rangle 
=\delta_{il}\delta_{jk}\, .
\label{twopoint}
\end{equation} 
Since we are not explicitly displaying the spacetime dependences, it is important to point out that all holomorphic operators are inserted at 
a specific spacetime event and all anti-holomorphic correlators are inserted at a second spacetime event. These correlators are called extremal
correlators; there are non-renormalization theorems protecting these correlators\cite{Lee:1998bxa,protected}.

When we talk about a half-BPS operator in what follows, we mean an operator built only from $Z$s. These operators will not break any further 
supersymmetries beyond those broken by the background itself. We obtain almost BPS operators by sprinkling $Y$s and $X$s in the operator.

\subsection{Super Yang-Mills Amplitudes}

The exact computation of multi-trace correlators at 
zero coupling is most easily carried out by expressing the multi-trace operators of interest in terms of Schur polynomials
\begin{equation}
\prod_{i}\Tr (Z^{n_i})=\sum_R\alpha_R\chi_{R}(Z),\qquad \prod_{j}\Tr ((Z^\dagger)^{m_j})=\sum_R\beta_R\chi_{R}(Z^\dagger ).
\label{pure}
\end{equation}
The coefficients $\alpha_R$ and $\beta_R$ appearing in the above expansion have no dependence on $N$. It will be useful first
to compute these correlators with a trivial background. A little Schur magic now gives (see Appendix D)
\begin{eqnarray}
{\cal A}(\{ n_i;m_j \},N)\equiv\left\langle \prod_{i,j}\Tr (Z^{n_i})\Tr ((Z^\dagger)^{m_j})\right\rangle &=&
\sum_{R,T}\alpha_R\beta_T\left\langle \chi_{R}(Z)\chi_{T}(Z^\dagger )\right\rangle\nonumber\\
&=&\sum_{R}\alpha_R\beta_R f_R\, .
\label{noback}
\end{eqnarray}
In this last equation $f_R$ is the product of the weights\footnote{Recall that the box in row $i$ and column $j$ has a weight
$N+j-i$.} (one for each box in the Young diagram) for Young diagram $R$. 

In what follows, $B$ denotes the rectangular Young diagram with $N$ rows and $M$ columns. We will often refer to $B$ as the
annulus background because the corresponding LLM geometry is obtained by taking the LLM boundary condition to be a black annulus 
on a white plane. The expectation value of an operator $O$ in background $B$ is given by
\begin{equation}
\left\langle O \right\rangle_B\equiv {\left\langle \chi_B(Z)\chi_B(Z^\dagger) O\right\rangle\over\left\langle \chi_B(Z)\chi_B(Z^\dagger)\right\rangle}\, .
\end{equation}
Our normalization is chosen so that the expectation value of the identity is 1. Recall that $B$ is a Young diagram with $N$ rows and $M$ columns.
The product rule satisfied by the Schur polynomials
can be viewed as a consequence of the fact that the Schur polynomials themselves are characters of $SU(N)$ and that (i) the character 
of a direct product of representations is just the product of the characters and (ii) the character of a given reducible representation is
equal to the sum of characters of the irreducible representations appearing in the reducible representation. In the product
$$\chi_R(Z)\chi_S(Z)=\sum_T g_{RST}\chi_T(Z),$$
the Littlewood-Richardson number $g_{RST}$ counts the number of times irreducible representation $T$ appears in the direct product of the
irreducible representations $R$ and $S$. Since we choose our background $B$ to be a Young diagram with $M$ columns and $N$ rows, it is a singlet
of $SU(N)$ and consequently we have the product shown in figure 1. 

\myfig{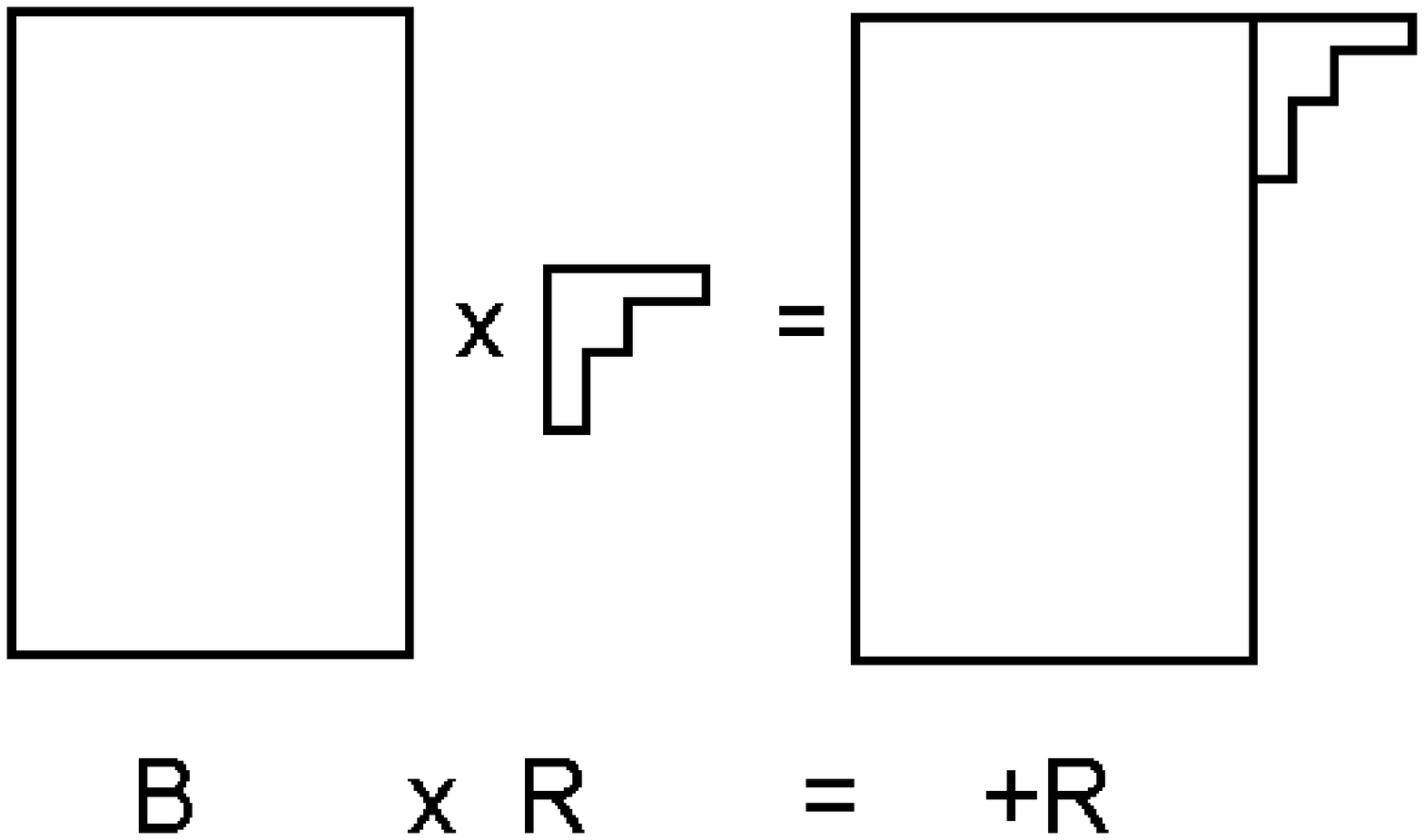}{6.5}{The product rule used to compute correlators in background $B$. This figure defines $+R$.}

Using this product rule, the correlator in the annulus background is
\begin{eqnarray}
{\cal A}_B(\{ n_i;m_j \},N)\equiv\left\langle \prod_{i,j}\Tr (Z^{n_i})\Tr ((Z^\dagger)^{m_j}\right\rangle_B &=&
\sum_{R,T}\alpha_R\beta_T 
{\left\langle \chi_{B}(Z)\chi_{R}(Z)\chi_{B}(Z^\dagger )\chi_{T}(Z^\dagger )\right\rangle\over f_B}\nonumber\\
&=& \sum_{R,T}\alpha_R\beta_T{\left\langle \chi_{+R}(Z)\chi_{+T}(Z^\dagger )\right\rangle\over f_B}\nonumber\\
&=&\sum_{R}\alpha_R\beta_R {f_{+R}\over f_B}\, .
\label{back}
\end{eqnarray}
Recall that $f_{+R}$ is the product of the weights appearing in Young diagram $+R$ and $f_B$ is the product of the weights appearing in
Young diagram $B$. All of the weights appearing in the product $f_B$ are repeated in the product $f_{+R}$ so that after canceling common 
factors $f_{+R}/f_B$ is simply equal to the product of the weights of the extra boxes stacked to the right of $B$ to form $+R$.
Thus, $f_{+R}/f_B$ is obtained from $f_R$ by replacing $N\to N+M$. Comparing (\ref{back}) and (\ref{noback}) we find
\begin{equation}
{\cal A}_B(\{ n_i;m_j \},N)={\cal A}(\{ n_i;m_j \},M+N).
\label{amplituderelation}
\end{equation}
We know that the correlator ${\cal A}(\{ n_i;m_j \},N)$ admits an expansion in ${1\over N}$; (\ref{amplituderelation}) tells us that 
${\cal A}_B(\{ n_i;m_j \},N)$ admits an expansion in ${1\over N+M}$. If we assume that $\sum_i n_i\sim O(1)$, we obtain correlators of 
operators that are dual to gravitons. Thus, for gravitons in the background $B$ ${1\over N+M}$ clearly plays the role of a loop expansion
parameter. {\sl Thus, we learn that there is a reorganization of the ${1\over N}$ expansion for these correlators in the
annulus background $B$ and further, that the new expansion parameter is ${1\over N+M}$.} Our relation (\ref{amplituderelation})
implies that the only effect of the background is to shift $N\to N+M$. 

One can easily check that this relation (\ref{amplituderelation}) 
is not a property of the full theory. Although a slight modification of (\ref{amplituderelation}) does allow us to relate the one point 
functions\footnote{Again computed in the theory with gauge group $U(N+M)$.} 
$\langle\Tr (Z^nZ^{\dagger\, n})\rangle$ and $\langle\Tr (Z^nZ^{\dagger\, n})\rangle_B$, we have not worked out a relation between
amplitudes in general. Our interest in the one point functions $\langle\Tr (Z^nZ^{\dagger\, n})\rangle_B$ is that they appear in the intermediate steps of
the computation of correlators of BMN-like probes of the annulus so that we manage to obtain a simple relation between the trivial background and the 
annulus background for both the BPS and near-BPS sectors of the theory.

The amplitudes ${\cal A}(\{ n_i;m_j \},M+N)$ are the amplitudes of a theory with a gauge group of rank $N+M$ and no background. In the LLM language, 
the boundary condition for this geometry is simply a black disk of radius $\sqrt{N+M}$. In our theory (with gauge group of rank $N$) the 
background $B$ corresponds to an annulus with inner radius $\sqrt{M}$ and outer 
radius $\sqrt{M+N}$. Thus, one way to interpret the relation (\ref{amplituderelation}) is that {\it any} of the half-BPS probes considered above
are unable to detect the hole in the middle of the annulus. We have not considered probes built using ${d\over dZ}$ instead of $Z$; these will 
detect the hole. Indeed, acting with traces of $Z$ on the background $\chi_B(Z)$ produces a new Schur polynomial that has extra boxes stacked
adjacent to the upper right hand corner of $B$ - this corner maps to the outer edge of the annulus defining the LLM boundary condition. Acting with
${d\over dZ}$ erodes corners from the lower right corner of $B$ - this corner maps to the inner edge of the annulus. See \cite{deMelloKoch} for further
details. Probes built using ${d\over dZ}$ instead of $Z$ are also half-BPS probes. 

We have related amplitudes in the gauge theory with a background of $M$ giant 
gravitons and gauge group $U(N)$ to amplitudes in the gauge theory with trivial background and gauge group $U(N+M)$. This is reminiscent
of the infrared duality proposed in \cite{Balasubramanian:2001dx} which exchanges the rank of the gauge group and the number of giant 
gravitons. In fact, (\ref{amplituderelation}) tells us that half-BPS correlators computed in the $U(N)$ gauge theory in the background of $M$
giant gravitons are exactly equal to the same half-BPS correlators computed in the $U(M)$ gauge theory in the background of $N$ giant gravitons.
Since these correlators are extremal and hence not renormalized, our computations give the value of these correlators in the deep infrared
limit of the gauge theory and thus seem to provide nontrivial support for the duality proposed in \cite{Balasubramanian:2001dx}. Note however, that 
correlators of operators built using ${d\over dZ}$ (which are also extremal) will not agree. This is not obviously in conflict with the proposed
duality of \cite{Balasubramanian:2001dx}. In \cite{Balasubramanian:2001dx} near extremal black holes are considered. These can be understood
as a condensate of giant gravitons\cite{Myers:2001aq} represented in ${\cal N}=4$ super Yang-Mills theory by a Schur polynomial corresponding
to a triangular Young diagram \cite{Balasubramanian:2005mg}. The LLM boundary condition for our background is a black annulus and ${d\over dZ}$ correlators
explore the inner edge of the annulus. The LLM boundary condition dual to the Schur polynomial with triangular Young diagram label is a gray disk; 
there is no inner edge. It would be interesting to see how many of our results for the black annulus boundary condition can be generalized
to the gray disk boundary condition. This generalization is nontrivial.

The main evidence given in \cite{Balasubramanian:2001dx} for the proposed duality was an exact correspondence between the gravitational entropy
formulae of small and large charge solutions. We can give arguments, in the free field theory, that suggest that the entropy of the state formed 
by $N_g$ condensed giant gravitons in the theory obtained by taking the near-horizon limit of $N$ D3-branes is equal to the entropy of the state 
formed by $N$ condensed giant gravitons in the theory obtained by taking the near-horizon limit of $N_g$ D3-branes. This provides further evidence
for the duality of \cite{Balasubramanian:2001dx} in a completely different region of parameter space to that probed by supergravity. According 
to\cite{Balasubramanian:2007bs} the near extremal black holes in AdS space can be obtained by attaching open string excitations to the 
Schur polynomials with triangular Young diagram labels. If we are very close to extremality, the number of open string excitations attached is
much smaller compared to the total number of fields in the operator. In this situation instead of attaching open strings by adding boxes to the 
original triangular label, it should be a good approximation to fix the tableau shape and replace some boxes (in arbitrary places on the Young 
diagram) with open strings. Assuming that the state obtained by deleting the open strings is again a typical state, i.e. again a triangle we
have something like the situation given in figure 2. To compute the entropy associated with these states, we need to count the number of different
ways of attaching the open string defects. This is given by the number of different ways we can pull the occupied boxes off the triangular Young
diagram. We don't know exactly how to compute this number. However, it is easy to argue that this number is invariant under flipping the
Young diagram so that the two states in figure 2 are swapped: this follows from two facts (i) the counting problem is constrained by the rule that
we can pull boxes off in any order as long as after each box is pulled off we continue to have a valid Young diagram\footnote{This is required because
removing each box must define a valid subduction. See\cite{Balasubramanian:2004nb,deMelloKoch:2007uu,deMelloKoch:2007uv,Bekker:2007ea}.} and (ii) a
shape that is (is not) a valid Young diagram before the flip is (is not) a valid Young diagram after the flip. Thus, the entropy of these two
states agree.
\myfig{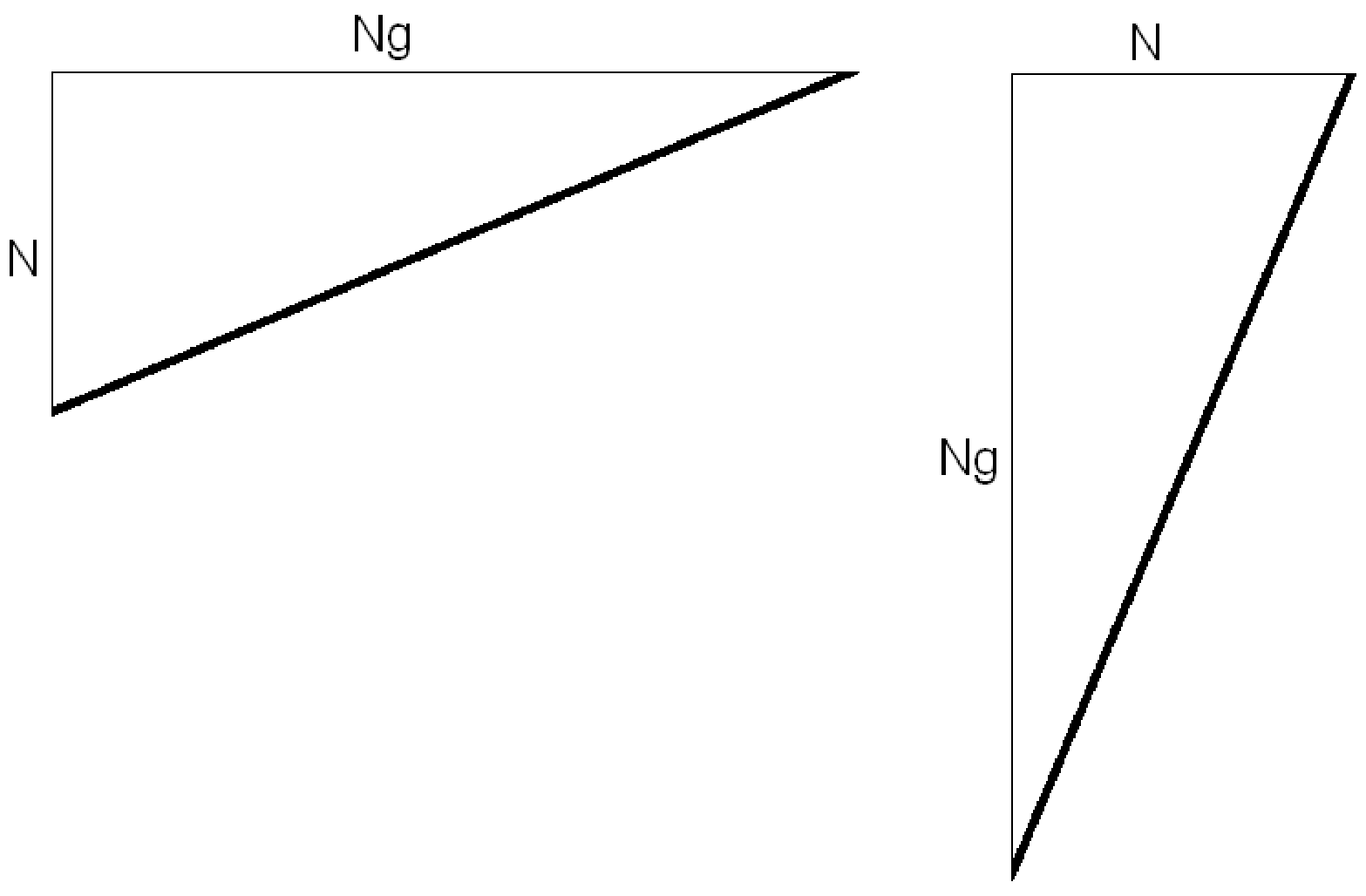}{6.5}{The typical state of $N_g$ condensed giant gravitons in the theory with gauge group $U(N)$ (on the left) and 
the typical state of $N$ condensed giant gravitons in the theory with gauge group $U(N_g)$ (on the right). The black stripe represents
the boxes occupied by open string defects.}

A few comments about the above flip of the Young diagram are in order.
The half-BPS sector of ${\cal N}=4$ super Yang-Mills theory can be mapped to the quantum Hall system with filling factor equal to one
\cite{Berenstein:2004hw,Ghodsi:2005ks}. In the half-BPS sector the above flip of the Young diagram is a $Z_2$ symmetry that exchanges
particles and holes\cite{Berenstein:2004hw,Ghodsi:2005ks,Alishahiha:2005fc}. If one employs a continuum (field theory) description of
the quantum Hall fluid, the resulting fluid is incompressible so that the corresponding continuum Lagrangian has a gauge invariance
under area preserving diffeomorphisms\cite{Susskind:2001fb}. Small fluctuations of the density, for a fluid of charged particles in
a background magnetic field, are well described by a Chern-Simons theory; the $U(1)$ gauge invariance is nothing but the area
preserving diffeomorphisms for the small fluctuations\cite{Susskind:2001fb}. The fluid description correctly captures the long distance
physics of the quantum Hall effect. It does not capture the fact that, since the fluid is described by $N$ electrons, it has an
intrinsic granular structure. One can capture this granular structure by using a Chern-Simons matrix model description\cite{Susskind:2001fb}.
In this Chern-Simons matrix model description the above $Z_2$ symmetry is nothing but level/rank duality of the 
Chern-Simons matrix model\cite{Lin:2005nh,Mosaffa:2006qk}.

One puzzling feature of the duality of \cite{Balasubramanian:2001dx} is the fact that it mixes giant gravitons and the branes (called background branes in
\cite{Balasubramanian:2001dx}) whose near horizon geometry is the AdS space we work in. There is a big difference between the background branes and the
giant gravitons. Indeed, in string theory the background branes carry a net RR-charge; giant gravitons carry no net charge - they are dipoles. In the dual
super Yang-Mills theory, changing the number of background branes changes the rank of the gauge group i.e. the number of fields we integrate over when
performing a path integral quantization. Changing the number of giant gravitons leaves the rank of the gauge group unchanged, but it does change the background
i.e. the integrand we use when performing a path integral quantization. If the duality of \cite{Balasubramanian:2001dx} is correct, we need to understand
how, in this case, changing the integrand has exactly the same effect as changing the number of
variables over which we integrate. Computing ${1\over 2}$-BPS correlators
in the annulus geometry gives a nice toy model in which to explore this issue. This is because we can reduce the whole problem to eigenvalue dynamics in zero
dimensions. Consider the computation of the correlator $\left\langle \Tr (Z)\Tr (Z^{\dagger})\right\rangle$ in the trivial vacuum of the $U(N+M)$ theory
($\Delta$ is usual the Van der Monde determinant)
\begin{equation}
\int \prod_{i=1}^{M+N} dz_i d\bar{z}_i \Delta (z)\Delta (\bar{z})\, \sum_{j=1}^{N+M} z_j\sum_{k=1}^{N+M}\bar{z}_k \, e^{-\sum_{i=1}^{N+M} z_i\bar{z}_i}\, .
\label{tivvac}
\end{equation}
The same correlator in the $U(N)$ theory with a background of $M$ giant gravitons is
\begin{equation}
\int \prod_{i=1}^{N} dz_i d\bar{z}_i \Delta (z)\Delta (\bar{z})\prod_{l=1}^N (z_l\bar{z}_l)^M
\, \sum_{j=1}^{N} z_j\sum_{k=1}^{N}\bar{z}_k \, e^{-\sum_{i=1}^{N} z_i\bar{z}_i}\, .
\label{gntvac}
\end{equation}
We already know that (\ref{tivvac}) and (\ref{gntvac}) give the same result. The reason why the two agree is now evident: in (\ref{tivvac}) we integrate over
an extra $M$ variables; these extra contributions add to give a larger result than that obtained for the same correlator in the $U(N)$ theory. In (\ref{gntvac})
the extra factor in the integrand implies that the integrand now peaks at larger values for the $|z_l|$; this again gives a larger result than that obtained for 
the same correlator in the $U(N)$ theory in the trivial vacuum. Our computation gives a simple picture of how, 
in this case, changing the integrand has exactly the same effect as changing the number of variables 
over which we integrate. It also suggests that the duality of \cite{Balasubramanian:2001dx} might be derived by starting with a suitable $U(N+N_g)$ theory and
(i) integrating out the degrees of freedom associated with $N_g$ colors to get a $U(N)$ gauge theory with a background given by a condensate of $N_g$ giant 
gravitons or (ii) integrating out the degrees of freedom associated with $N$ colors to get a $U(N_g)$ gauge theory with a background given by a condensate of 
$N$ giant gravitons.

\subsection{Supergravity Amplitudes}

According to the AdS/CFT correspondence, correlation functions can be computed in the strong coupling limit of ${\cal N}=4$ super Yang-Mills theory
using the dual hologram, which is type IIB supergravity. The paper \cite{Skenderis:2007yb}, has given a powerful general approach to holography
in the LLM backgrounds, generalizing and extending the Coulomb branch analysis of \cite{Skenderis:2006di,Klebanov:1999tb}. The formalism
given in \cite{Skenderis:2007yb} is an application
of the general method of \cite{Skenderis:2006uy} which employs the method of holographic renormalization\cite{Skenderis:2002wp}. An important result of
\cite{Skenderis:2007yb} was the demonstration that the asymptotics of the LLM solutions correctly encode the vacuum expectation values of all
single trace ${1\over 2}$ BPS operators to the leading order in the large $N$ expansion. {\sl Can we reproduce the results of the last section using
holography in the LLM background?}

One case of interest to us in this article is that of graviton three point functions, in the background created by a heavy operator. We have computed these
correlation functions in the free field theory; supergravity will reproduce these correlators in the strong coupling limit. We will now argue that it is 
natural to expect that these two results will agree since we can argue that there is a non-renormalization theorem protecting the three point functions 
of interest to us. The logic of this argument is very similar to the argument of \cite{Skenderis:2007yb} which argued that the generic one point function 
in the LLM background is protected. The three point functions we are interested in 
$$ \left\langle {\rm Tr} (Z^n){\rm Tr} (Z^m){\rm Tr} (Z^{\dagger\, m+n})\right\rangle_B $$
can of course, also be understood as a ratio of (higher point) correlators in the original trivial background
$$ {\left\langle \chi_B(Z)\chi_B(Z^\dagger ){\rm Tr} (Z^n){\rm Tr} (Z^m){\rm Tr} (Z^{\dagger\, m+n})\right\rangle\over
\left\langle \chi_B(Z)\chi_B(Z^\dagger ) \right\rangle}\, . $$
The correlators appearing in the above expressions are extremal correlators in the language of \cite{D'Hoker:1999ea}.
The computations of \cite{D'Hoker:1999ea} show that at the leading order in large $N$, the extremal correlators take the same
value at large 't Hooft coupling as in the free field theory and hence it is natural to expect that all extremal correlators
are not renormalized\footnote{Note that the supergravity result suggests that the planar contribution is not renormalized.
Here we are using the stronger conjecture \cite{D'Hoker:1999ea} which claims the non-renormalization for any $N$. Our results
provide further evidence for this stronger conjecture.}.

To extract three point functions using the methods of \cite{Skenderis:2007yb}, we would need to solve the equations of motion, to quadratic order,
around the LLM solution. We have not managed to do this. For the case of extremal correlators, there is some room for optimism: the analysis of 
\cite{D'Hoker:1999ea} argued that for extremal correlators, the bulk cubic supergravity coupling vanishes and the entire contribution comes from 
a boundary term. The LLM geometries are asymptotically AdS so that one might have hoped that there was a simple explanation of the results of the
previous section. We have not found one. Note also that the fact that the bulk cubic supergravity coupling vanishes in the AdS$_5$ background need not
imply that it continues to vanish in the LLM background.

Rather than pursuing the computation of three (and higher) point functions directly, we have found it simpler to relate the three point functions 
we'd like to compute to one point functions, since these have already been obtained in \cite{Skenderis:2007yb}. For concreteness, consider
the three point function ($B$ is again the annulus diagram)
$$ \left\langle {\rm Tr}(Z^2){\rm Tr}(Z^2){\rm Tr}(Z^{\dagger\, 4})\right\rangle_B $$
which can also be written as the one point function of ${\rm Tr}(Z^2)$ in the normalized state
$$ |\Phi\rangle = {\cal N}(\Tr (Z^2)+\Tr (Z^4))\chi_B(Z)|0\rangle \, .$$
${\cal N}$ is a normalization factor. Now, summing only planar diagrams
$$ \left\langle {\rm Tr}(Z^2){\rm Tr}(Z^2){\rm Tr}(Z^{\dagger\, 4})\right\rangle = 16 N^3 $$
so that according to our result of the previous section
$$ \left\langle {\rm Tr}(Z^2){\rm Tr}(Z^2){\rm Tr}(Z^{\dagger\, 4})\right\rangle_B = 16(N+M)^3\, .$$
The supergravity one point function is computed with a normalized state $|\Phi\rangle$; the above three point
function is computed with a different normalization. Noting that (we are again using results from the previous 
section)\footnote{Note that since $|\Phi\rangle$ has two additive components the one-point function has additional
cross terms; these however evaluate to zero leaving only the term we wish to compute.}
$$ 
\left\langle(\Tr (Z^2)+\Tr (Z^4))(\Tr (Z^{\dagger\, 2})+\Tr (Z^{\dagger\, 4}))\chi_B(Z) \chi_B(Z^\dagger)\right\rangle
$$
$$
=\left[ 4(N+M)^4+22(N+M)^2\right]f_B=4(N+M)^4 f_B\left( 1+O((N+M)^{-2})\right),
$$
we easily find
$$\langle \Phi |{\rm Tr}(Z^2)|\Phi\rangle = {4\over N+M}\, .$$ 
This is the result we would like to reproduce from supergravity.

We will summarize the main steps involved in extracting correlators of local operators in the boundary gauge theory, 
from the ten dimensional asymptotically AdS$_5\times$S$^5$ solutions of Type IIB supergravity. For all the details
and further references consult \cite{Skenderis:2007yb}. Given an asymptotically AdS$_5\times$S$^5$ solution of Type IIB supergravity, 
one must first systematically reduce the ten dimensional solution to a solution of five dimensional gravity coupled to an infinite 
number of five dimensional fields. If this reduction is performed in terms of gauge invariant variables (as it is in \cite{Skenderis:2007yb}),
it does not matter in which gauge the original solution is supplied. This reduction is then supplemented with a non-linear field redefinition,
performed in such a way that the equations of motion in terms of the redefined fields can be obtained by minimizing a local five dimensional 
action. One can then apply the usual holographic rules to compute the correlation function of boundary operators from 
the bulk five dimensional description. To obtain the renormalized correlation functions in the gauge theory one must appropriately renormalize 
the bulk gravitational action using holographic renormalization. The supergravity field that couples to the boundary operator that we are 
interested in (${\rm Tr}(Z^2)$) is a mixture of (components of the) metric on the S$^5$ and the four form RR potential on the $S^5$. It is
denoted by $S^{22}$ in \cite{Skenderis:2007yb}. The one point function we want can now be obtained by variation of the renormalized on 
shell supergravity action with respect to the boundary value of $S^{22}$. The result is \cite{Skenderis:2007yb}\footnote{This differs by
a factor of $\sqrt{2}\pi^2$ from formula (3.60) of \cite{Skenderis:2007yb} because we are using a different normalization. Further, since we have
dropped all spacetime dependence, we have also dropped a factor of $e^{2it}$.}
$$ \left\langle {\cal O}_{S^{22}}\right\rangle = N^2 \int r^3\rho\, e^{2i\phi} dr d\phi\, ,$$
where $\rho$ is the boundary condition for the LLM geometry. This boundary condition is most easily determined by exploiting the free
fermion description of the half-BPS sector of ${\cal N}=4$ super Yang-Mills theory\cite{Corley:2001zk,Berenstein:2004kk}. To make the transition
to the free fermion description, rewrite the state $|\Phi\rangle$ in terms of Schur polynomials. This is easily accomplished using
$$ {\rm Tr} (Z^2)=\chi_{\tiny \yng(2)}(Z)-\chi_{\tiny \yng(1,1)}(Z)\, ,$$
$$ {\rm Tr} (Z^4)=\chi_{\tiny \yng(4)}(Z)-\chi_{\tiny \yng(3,1)}(Z)+\chi_{\tiny \yng(2,1,1)}(Z)-\chi_{\tiny \yng(1,1,1,1)}(Z)\, ,$$
and the product rule for Schur polynomials. The Schur polynomials correspond to energy eigenkets of the free fermions. The energies of the
free fermions in the state are
$$ E_i=\lambda_i+N-i+1 \, \qquad i=1,...,N,$$
where $\lambda_i$ is the number of boxes in the $i$th row of the Young diagram label for the Schur polynomial. Using this interpretation it is 
straightforward to obtain the fermion phase space density. The density we obtain in this way is
$$\rho=\rho_1+\cos (2\phi)\rho_2 \, ,$$
where
$$
\rho_2={1\over 2\pi (N+M)^4}{e^{-Nr^2}\over (N+M-1)!}\left[
(Nr^2)^{N+M+2}+(N+M-1)(Nr^2)^{N+M+1}\right.
$$
$$
-(N+M)(N+M+1)(N+M-1)(N+M-2)(Nr^2)^{N+M-2}
$$
$$
\left. -(N+M)(N+M-1)^2(N+M-2)(N+M-3)(Nr^2)^{N+M-3}
\right]\, .
$$
We will not need $\rho_1$ in what follows.
As already noted in \cite{Skenderis:2007yb}, the state $|\Phi\rangle$ which is a superposition 
of a small number (for us, 6) of Schur polynomials, does
not give rise to a smooth supergravity solution. Indeed, for a regular supergravity solution, $\pi\rho$ should only take the values $\{ 0,1\}$;
this is not the case for the above $\rho$. It is now straightforward to check that
$$ \left\langle {\cal O}_{S^{22}}\right\rangle = {4\over N+M} $$
which is the result we wanted to demonstrate.

\subsection{Multi Ring Backgrounds}

Using the results of \cite{deMelloKoch,Koch:2008cm} we can generalize our results for the annulus to a
general multi ring LLM geometry. The Young diagrams of these geometries are not a rectangle, but rather are shapes with more
than 4 edges and all edges have a length of order $N$ boxes. In a geometry with $m$ thick rings, local 
graviton operators were defined in \cite{deMelloKoch}. When taking a product of the Young diagram of the 
background ($B_{ring}$) with the Young diagram of the probe ($R$), one removes boxes from the probe Young diagram 
and adds them, at all possible positions, to $B_{ring}$ (respecting the usual rules for multiplying Young diagrams). 
The local graviton operators are defined by giving their product rule: the boxes removed from $R$ can only be
added at a specific location on $B_{ring}$. This definition is precise enough to allow the computation of
correlators; see \cite{deMelloKoch} for more details. We can also give a rough constructive
definition of the local graviton operators: since we consider a boundary condition with $m$ thick rings, the
eigenvalue distribution in the dual matrix model will split into $m$ well separated clumps.
In the limit that we expect a classical geometry to emerge (large $N$ and large 't Hooft coupling) the 
off diagonal modes connecting these $m$ subsectors will be very heavy and decouple. We expect that,
when studying almost BPS states, the effect of these modes on the dynamics can be neglected. There is
no reason to neglect off diagonal modes connecting eigenvalues in the same sector. $Z$ can thus
be replaced by a block diagonal matrix with $m$ blocks. If clump $i$ contains $N_i$ eigenvalues it 
corresponds to an $N_i\times N_i$ block. To construct a local graviton operator, do not use the full
matrix $Z$, but rather use one of the blocks $Z_i$.

It is now straight forward to verify that, if ring $i$ has outer radius $\sqrt{N+M_i}$ then we again get an exact relation
\begin{equation}
{\cal A}_B(\{ n_i;m_j \},N)={\cal A}(\{ n_i;m_j \},M_i+N),
\end{equation}
for correlators ${\cal A}_B(\{ n_i;m_j \},N)$ of gravitons localized at edge $i$.
Thus, the gravitons at the edge of each ring have their own expansion parameter equal to ${1\over N+M_i}$.

Finally, consider a background obtained by exciting $k_2$ sphere giants; each giant is assumed to carry a momentum $k_3<N$. The corresponding
LLM geometry will be a black disk (of area $N-k_3$) surrounded by a white annulus of area $k_2$ which is itself surrounded by a black
annulus of area $k_3$ (this boundary condition is shown in figure 3 of \cite{Ghodsi:2005ks}). From the results of this section we know that
excitations of the central black droplet could be described using either a $U(N)$ theory with a background of $k_2$ sphere giants or using
a $U(N-k_3)$ theory. This is closely related to remarks made in \cite{Ghodsi:2005ks}, arrived at using a totally different approach.

\subsection{Backgrounds with $>1$ Charges}

In this section we would like to study the zero coupling limit of backgrounds made from two matrices $Z$ and $Y$. There are
a number of bases which we can employ for this study. The basis described in \cite{Brown:2007xh} builds operators with definite 
quantum numbers for any desired global symmetry groups acting on the matrices. The basis of \cite{Kimura:2007wy} uses the Brauer 
algebra to build correlators involving $Z$ and $Z^\dagger$; this basis seems to be the most natural for exploring brane/anti-brane systems.
The basis of \cite{Bhattacharyya:2008rb} most directly allows one to consider open string 
excitations\cite{Balasubramanian:2002sa,Balasubramanian:2004nb,deMelloKoch:2007uu,deMelloKoch:2007uv,Bekker:2007ea} of the operator; this 
is the restricted Schur basis. These three bases do not coincide and a detailed link between them has been discussed in \cite{Collins:2008gc}.
Casimirs that distinguish these bases have been given in \cite{Kimura:2008ac}. Finding a spacetime interpretation of these Casimirs 
is a promising approach towards understanding the dual description of these bases.
Although all three bases diagonalize the two point functions in the free field theory limit, computing anomalous dimensions is not
yet possible\footnote{For some progress see \cite{deMelloKoch:2007uv,Bekker:2007ea,Brown:2008rs}. In particular, these articles show that at one
loop, operators can only mix if their labels differ by the location of a single box; thus, it seems that the operator mixing should have a nice
simple description. This is proved in the restricted Schur basis in \cite{deMelloKoch:2007uv,Bekker:2007ea} and in the basis of \cite{Brown:2007xh}
in \cite{Brown:2008rs}.}. In particular, we do not know how to extract the multi-matrix BPS operators from these bases.

In \cite{Bhattacharyya:2008rc} a particularly simple set of restricted Schur polynomials were identified; these are the operators that
we will focus on in this section. The operator is built using $NM_1$ $Z$ fields and $NM_2$ $Y$ fields. The restricted Schur polynomial
has two labels $\chi_{R,(r_1,r_2)}$. $R$ is an irreducible representation of $S_{N(M_1+M_2)}$. We take $R$ to be a Young diagram with 
$M_1+M_2$ columns and $N$ rows. $(r_1,r_2)$ is an irreducible representation of the $S_{NM_1}\times S_{NM_2}$ subgroup. We take $r_1$
to be a Young diagram with $M_1$ columns and $N$ rows and $r_2$ to be a Young diagram with $M_2$ columns and $N$ rows. We take both
$M_1$ and $M_2$ to be $O(N)$. For this particular restricted Schur polynomial we have
$$\chi_{R,(r_1,r_2)}(Z,Y)={{\rm hooks}_{R} \over {\rm hooks}_{r_1}\, {\rm hooks}_{r_2}}\chi_{r_1}(Z)\chi_{r_2}(Y)\, ,$$
where $\chi_{r_1}(Z)$ and $\chi_{r_2}(Y)$ are standard Schur polynomials. The simplicity of this background is a consequence of the fact that 
the above operator factorizes. These operators are not BPS or even near-BPS. Indeed, in appendix C we compute the anomalous dimension of these
operators, which is given by
$$\Delta = N(M_1+M_2)+4\lambda M_1 M_2\, , \qquad \lambda = Ng_{YM}^2. $$
Thus, the results of this section can not be extrapolated to strong coupling. It is still interesting to ask if, at weak coupling, we can
reorganize the usual ${1\over N}$ expansion.

After normalizing so that the identity has an expectation value of 1, we find
$$
\left\langle O\right\rangle_{(r_1,r_2)}=
{\left\langle\chi_{r_1}(Z)\chi_{r_1}(Z^\dagger)\chi_{r_2}(Y)\chi_{r_2}(Y^\dagger)O\right\rangle\over
\left\langle\chi_{r_1}(Z)\chi_{r_1}(Z^\dagger)\chi_{r_2}(Y)\chi_{r_2}(Y^\dagger)\right\rangle}\, .
$$
Using exactly the same methods as were used in section 2.1, we find that the complete effect of the background on correlators of operators
built using only $Z$s with operators built using only $Z^\dagger$s is to replace $N\to N+M_1$. In addition, we find that the complete effect 
of the background on correlators of operators built using only $Y$s with operators built using only $Y^\dagger$s is to replace $N\to N+M_2$.
Thus, these two types of observables admit a different reorganization of the $1/N$ expansion; the small parameter of the two expansions
is not the same.

\subsection{A Generalization of the BMN Limit for Two Charge Backgrounds}

The AdS/CFT correspondence forces us to take the limit of large 't Hooft coupling $\lambda$, if 
we are to suppress curvature corrections in the string theory. It
is thus a strong/weak coupling duality. This naive observation is evaded by the remarkable result of BMN\cite{Berenstein:2002jq}: the expansion
parameter of the super Yang-Mills theory may be suppressed by large quantum numbers for ``almost BPS operators'' \cite{Berenstein:2003ah}. The
existence of double scaling BMN-like limits provides a very rich class of tractable problems. In this section we would like
to explore the proposed BMN-like sectors discovered in \cite{Fareghbal:2008ar}\footnote{We thank Shahin Sheikh-Jabbari for extremely useful
correspondence, which lead to the results of this subsection.}. The work \cite{Fareghbal:2008ar} itself was an extension of \cite{Balasubramanian:2007bs}.
The existence of these limits may be very relevant to understanding
the gauge theory description of near-extremal charged black holes in AdS$_5$. Using the technology we have developed, it should be possible
to study these limits.

We use the same two-charge background that we used in the previous subsection. We will however take $M_1=O(\sqrt{N})$ and $M_2=O(\sqrt{N})$ with
the 't Hooft coupling $\lambda$ large but fixed\footnote{This is not quite the same limit as that described in \cite{Fareghbal:2008ar}; in
\cite{Fareghbal:2008ar} the limit with $g_{YM}^2$ fixed is considered.}. In this case, the charges $J_1= NM_1\sim O(N^{3\over 2})$ and
$J_2= NM_1\sim O(N^{3\over 2})$, whilst the difference 
$$ \Delta -J_1-J_2 = 4\lambda M_1 M_2\sim N $$
so that\footnote{We kept the 't Hooft coupling fixed in order to obtain this scaling for $\eta$, which matches precisely what one obtains
in the BMN limit.}
$$ \eta\equiv {\Delta -J_1-J_2 \over J_1+J_2}\sim N^{-{1\over 2}}\to 0\, .$$
Our background (plus open string excitations which do not modify this scaling) is a near BPS operator, in the language of \cite{Berenstein:2003ah}.
These results are in perfect agreement with the supergravity arguments of \cite{Fareghbal:2008ar} - the strong and weak coupling results match. 
The operator we have obtained is a ``near ${1\over 4}$ BPS'' state. The fact that the anomalous dimension is proportional to $M_1 M_2$ strongly
suggests that it arises from the dynamics of the open strings stretching from the stack of $M_1$ giants (described by $\chi_{r_1}(Z)$) to the 
stack of $M_2$ giants (described by $\chi_{r_2}(Y)$).

Let us now recall a few well known facts. 't Hooft\cite{'tHooft:1973jz} has already made the beautiful observation that the perturbative 
expansion of a matrix model can be written in the form
\begin{equation}
\sum_{g=0}^\infty N^{2-2g} f_g(\lambda )\, ,
\label{expand}
\end{equation}
where $f_g (\lambda )$ is a polynomial in the 't Hooft coupling - there is no additional $N$ dependence apart from the multiplying factor of
$N^{2-2g}$. Here, ${1\over N}$ plays the role of a genus counting parameter.
In the BMN set up, one learnt that there is a new effective genus counting parameter $g_2={J^2\over N}$ replacing ${1\over N}$, by studying the two 
point functions of BMN loops (each has angular momentum $J$). In terms of this effective genus counting parameter, the effective 't Hooft coupling 
is $g_{YM}^2/g_2=g_{YM}^2N/J^2$. Can we repeat this analysis for the operators considered here? Towards this end, we set $M_1=M_2=M$ and compute the 
two point function
$$ I_2=\left\langle \chi_{r_1}(Z)^\dagger \chi_{r_2}(Y)^\dagger \chi_{r_1}(Z)\chi_{r_2}(Y)\right\rangle =\left(
{G_2 (N+M+1)\over G_2 (N+1)G_2 (M+1)}\right)^2 \, ,$$
where $G_2 (n+1)$ is the Barnes function defined by ($\Gamma (z)$ is the Gamma function)
$$ G_2 (z+1)=\Gamma (z)G_2 (z)\, ,\qquad G_2 (n+1)=\prod_{k=1}^{n-1} k!\, .$$
We will now set $N=\alpha M^2$ where $\alpha$ is a number of $O(1)$ in the large $N$ limit.
The Barnes function has the asymptotic expansion
$$ \log G_2 (N+1)={N^2\over 2}\log (N)-{1\over 12}\log (N)-{3\over 4}N^2 +{N\over 2}\log (2\pi) +\zeta'(-1)
+\sum_{g=2}^\infty {B_{2g}\over 2g(2g-2)N^{2g-2}}\, ,$$
where $B_{2g}$ are the Bernoulli numbers and $\zeta'(-1)$ is the derivative of the Riemann zeta function evaluated at -1.
Using this asymptotic expansion, it is clear that $I_2$ can be split into a product $F_{\rm non-pert} F_{\rm pert}$, where
$F_{\rm non-pert}$ is a non-perturbative piece that can not be expanded in ${1\over M}$ and $F_{\rm pert}$ is a factor that
does admit an expansion in ${1\over M}$\footnote{This factorization is not unique; it will be fixed by the physics of the problem.
We do not yet know how to do this.}. This suggests that we should identify $g_2={1\over M}\sim {1\over\sqrt{N}}$. In
the BMN case only even powers of $g_2$ appear in the genus expansion; here we can not be sure if only even powers of
$g_2$ appear. This depends on precisely how we factorize $I_2$ into $F_{\rm non-pert} F_{\rm pert}$. Using this genus
expansion, we would expect an effective 't Hooft coupling\footnote{With this scaling, which is different to what we had above,
$\eta$ does not scale with $N$, but can be made arbitrarily small.}
$$\tilde{\lambda}=g_{YM}^2 M={1\over \sqrt{\alpha}} g_{YM}^2\sqrt{N}. $$
Keeping this effective 't Hooft coupling fixed but arbitrarily small,
one finds
$$ \Delta= 2NM\left(1 + 2\tilde{\lambda}\right)=2\alpha M^3 \left(1 + 2\tilde{\lambda}\right) \, .$$
This looks like a polynomial is $\tilde{\lambda}$ times some power of $M$. Clearly, in view of (\ref{expand}),
$\tilde{\lambda}$ is indeed the correct 't Hooft coupling. To properly specify our state of intersecting giants, we need to specify
the three Young diagrams which would label the restricted Schur polynomial of the giant system. Clearly, with our simple choice it seems
that we have correctly obtained the correct Yang-Mills operator to describe the decoupling limits of \cite{Balasubramanian:2007bs,Fareghbal:2008ar}
to one loop. It would be interesting to compute higher loop corrections and verify that one does indeed continue to obtain a polynomial 
in $\tilde{\lambda}$ times $M^3$. If extra $M$ dependence does appear, it might be a sign that our simple guess for the background needs to be
corrected at $O(\tilde{\lambda})$. This is not completely unexpected: these operators are not protected so one would
expect corrections in general. Finally, holding $\tilde{\lambda}$ fixed we find the 't Hooft
coupling $\lambda=g_{YM}^2 N=\tilde{\lambda}\sqrt{\alpha N}$ goes to infinity. Thus, even though we have large $\lambda$ the 
one loop correction can be made arbitrarily small. We are thus optimistic that,
similar to what happens in the BMN limit, it will be possible to compare perturbative field theory results to results obtained
from the dual gravitational description.

The genus counting parameter $g_2={1\over M}$ and the effective 't Hooft coupling $\tilde{\lambda}=g_{YM}^2 M$ are exactly the parameters
we would expect for the description of the large $M$ limit of a $U(M)$ gauge theory. This is rather natural: we have on the order of $M$ giant
gravitons whose worldvolume theory at low energy will be a $U(M)$ gauge theory. This provides strong field theory evidence that the sector of 
the theory identified in \cite{Balasubramanian:2007bs,Fareghbal:2008ar} is indeed captured by the dynamics of open string defects distributed
on a boundstate of intersecting giant gravitons. Its natural to think that the decoupling limits of \cite{Balasubramanian:2007bs,Fareghbal:2008ar} capture 
the decoupled low energy world volume theory of the intersecting giant gravitons in the same way that Maldacena's limit\cite{Maldacena:1997re} captures the 
decoupled low energy world volume theory of $N$ D3 branes. Our explicit computations show that the low energy world volume theory of the giant 
gravitons is weakly coupled even when the original ${\cal N}=4$ super Yang-Mills theory is strongly coupled.

\section{Beyond the Half-BPS Sector}

In this section we would like to determine the sector of the theory in which our reorganization of perturbation theory 
is valid. First, the near BPS operators we are interested in are BMN-like loops
$${\cal O}(\{ n\})= \Tr (YZ^{n_1}YZ^{n_2}YZ^{n_3}Y\cdots Y Z^{n_L}).$$
To compute the two point correlator $\left\langle {\cal O}(\{ n\}){\cal O}^\dagger (\{ n\})\right\rangle_B$ we start by contracting the
$Y$ fields planarly. After performing the $Y$ contractions we need to compute
\begin{equation}
\left\langle \prod_i \Tr (Z^{n_i}Z^{\dagger \, n_i})\right\rangle_B \, . 
\label{onepoint}
\end{equation}
We need to verify that the above amplitude admits an expansion with parameter ${1\over (N+M)}$ to demonstrate that our reorganization
of perturbation theory does indeed apply to the half-BPS and almost BPS sectors. 

We will demonstrate that a relation very similar to (\ref{amplituderelation}) holds for the leading contribution to the
one point functions (\ref{onepoint}). We will then derive an exact relation. The first property we use is factorization (valid to leading order
and if $\sum_i n_i\sim O(1)$)
$$ \left\langle \prod_i \Tr (Z^{n_i}Z^{\dagger \, n_i})\right\rangle_B 
= \prod_i \left\langle \Tr (Z^{n_i}Z^{\dagger \, n_i})\right\rangle_B \, . $$
Factorization in the annulus background has been discussed in \cite{deMelloKoch} and in Appendix A. Thus, we need only consider
$$
\left\langle \Tr (Z^{n}Z^{\dagger \, n})\right\rangle_B = \left\langle \Tr (Z^{n}Y) \Tr(Z^{\dagger \, n}Y^\dagger)\right\rangle_B\, . 
$$

This computation can be completed in exactly the same way as the computation we tackled in section 2.1. We start by writing the loop 
of interest in terms of restricted Schur polynomials
\begin{equation}
{\cal O}(n)= \Tr (Z^{n}Y) =\sum_{(R,R')}\alpha_{(R,R')}\chi_{(R,R')}(Z,Y).
\label{notpure}
\end{equation}
$R$ is a Young diagram with $n+1$ boxes and $R'$ a Young diagram with $n$ boxes. The expansion coefficients $\alpha_{R,R'}$ are
independent of $N$. The two point function of restricted Schur polynomials has been computed in \cite{deMelloKoch:2007uu,Bhattacharyya:2008rb}. 
Using these results, we find
$$
{\cal C}(n,N)=\left\langle {\cal O}(n){\cal O}^\dagger(n)\right\rangle =\sum_{(R,R')}\alpha_{R,R'}^2 
{{\rm hooks}_R\over {\rm hooks}_{R'}}f_R\, .
$$
The factor ${\rm hooks}_R/{\rm hooks}_{R'}$ does not depend on $N$.
To compute these amplitudes in the annulus background, we need the analog of the product rule given in Fig. 1. The product rule for restricted
Schur polynomials\cite{Bhattacharyya:2008rc} was used in \cite{Koch:2008cm} to compute precisely the restricted Littlewood-Richardson number
needed here. Again, only a single term enters in the product
$$ \chi_B(Z)\chi_{(R,R')}(Z,Y)=g_{B\,(R,R')\, (+R,+R')}\,\, \chi_{(+R,+R')}(Z,Y)
.$$
$+R'$ is obtained from $+R$ by dropping a single box.
This restricted Littlewood-Richardson number was computed in \cite{Koch:2008cm} from its
definition; this involves a summation over restricted characters. There is however, a much simpler way to obtain this result. Once one has established
the form $ \chi_B(Z)\chi_{(R,R')}(Z,Y)=g_{\tiny B\,(R,R')\, (+R,+R')}\,\, \chi_{(+R,+R')}(Z,Y)$, reducing both sides with respect to $Y$ gives\footnote{To
reduce with respect to $Y$ we take the derivative $\Tr {d\over dY}$. A formula for the reduction of Schur polynomials is given in \cite{deMelloKoch:2004ws};
a formula for the reduction of restricted Schur polynomials is given in \cite{deMelloKoch:2007uu}.}
$$ c_{R,R'}\chi_B (Z)\chi_{R'}(Z)=c_{R,R'}\chi_{+R'}(Z)=c_{+R,+R'}g_{\tiny B\,(R,R')\, (+R,+R')}\,\, \chi_{+R'}(Z),$$
where we have used the product rule of section 2.1: $\chi_{B}(Z)\chi_{R'}(Z)=\chi_{+R'}(Z)$ and where $c_{T,T'}$ is the weight of the box that
must be dropped from $T$ to obtain $T'$. Thus,
$$g_{\tiny B\,(R,R')\, (+R,+R')}={c_{R,R'}\over c_{+R,+R'}}. $$
This formula is exact. Although it is possible to compute things exactly, we will also make good use of the leading large $N+M$ version of our results.
For this reason we start with a large $N+M$ analysis and then give exact results.

\noindent
{\bf Large $N$ Analysis:}
To leading order in large $N+M$ we have $c_{R,R'}=N$ and $c_{+R,+R'}=N+M$ which reproduces the known result\cite{Koch:2008cm}. It is now straight forward
to find
\begin{eqnarray}
{\cal C}_B(n,N) &=&\left\langle {\cal O}(n){\cal O}^\dagger(n)\right\rangle_B\nonumber \\
&=&\sum_{(R,R')}\alpha_{R,R'}^2\left({N\over N+M}\right)^2 
{{\rm hooks}_{+R}\over {\rm hooks}_{+R'}}{f_{+R}\over f_B}\nonumber\\
&=& \left({N\over N+M}\right)\sum_{(R,R')}\alpha_{R,R'}^2 
{{\rm hooks}_{R}\over {\rm hooks}_{R'}}{f_{+R}\over f_B}\, .
\end{eqnarray}
Arguing exactly as we did in section 2.1 we find
\begin{equation}
{\cal C}_B(n,N)=\left({N\over N+M}\right) {\cal C} (n,N+M)\, .
\label{secondrelation}
\end{equation} 
This is very similar to, but not quite the same as (\ref{amplituderelation}). Note also that
(\ref{secondrelation}) was only derived at leading order; (\ref{amplituderelation}) is however exact.
Some insight into these relations can be obtained by noticing that at leading order in $N$ we have
$$ \left\langle \Tr (Z^n)\Tr (Z^{\dagger\, n})\right\rangle = N^n,\qquad
\left\langle \Tr (Z^n Z^{\dagger\, n})\right\rangle =N^{n+1}.$$
Thus, as long as we restrict ourselves to the leading order, (\ref{amplituderelation}) and (\ref{secondrelation}) 
can be restated as follows: consider an operator $O(Z,Z^\dagger)$ which is a product of factors of the form 
(\ref{pure}) or (\ref{notpure}). Then
\begin{equation}
\left\langle O(Z,Z^\dagger)\right\rangle_B =\left\langle O\left(\sqrt{N+M\over N}Z,\sqrt{N+M\over N}Z^\dagger\right)\right\rangle\, .
\label{rescale}
\end{equation}
{\sl At leading order, the only effect of the background is to rescale $Z$.} This observation will be useful below.

\noindent
{\bf Exact Analysis:}
The box $c_{RR'}$ has a weight $N+p$ where $p$ is a number of $O(1)$. Then 
$$g_{\tiny B\,(R,R')\, (+R,+R')}={N+p\over N+M+p},\qquad {{\rm hooks}_{+R}\over {\rm hooks}_{+R'}}
={N+M+p\over N+p}{{\rm hooks}_{R}\over {\rm hooks}_{R'}}\, ,$$
so that
$$ \left\langle \Tr (Z^n Y)\Tr (Z^{\dagger\, n}Y^\dagger )\right\rangle_B =\sum_{(R,R')}
{N+p\over N+M+p}{{\rm hooks}_{R}\over {\rm hooks}_{R'}}{1\over (n+1)^2}{f_{+R}\over f_B}$$
$$=\sum_{(R,R')}
{{\rm hooks}_{R}\over {\rm hooks}_{R'}}{1\over (n+1)^2}{f_{+R}\over f_B}
-M\sum_{(R,R')} {1\over N+M+p}{{\rm hooks}_{R}\over {\rm hooks}_{R'}}{1\over (n+1)^2}{f_{+R}\over f_B}$$
$$=\sum_{(R,R')}
{{\rm hooks}_{R}\over {\rm hooks}_{R'}}{1\over (n+1)^2}{f_{+R}\over f_B}
-M\sum_{(R,R')} {{\rm hooks}_{R}\over {\rm hooks}_{R'}}{1\over (n+1)^2}{f_{+R'}\over f_B}\, .$$
In the above sums, $R$ runs over all hook Young diagrams with $n+1$ boxes (diagrams with at most one column containing more than one box; see
\cite{Koch:2008cm}) and $R'$ runs over all possible ways of removing a box from the hook (there are usually 2 possible ways).
The first term in this last expression is nothing but ${\cal C}(n,N+M)$. Now, it is a simple matter to verify that
\begin{equation}
\sum_{R} {{\rm hooks}_{R}\over {\rm hooks}_{R'}}{1\over (n+1)^2} =
\sum_{R} {d_{R'}\over d_{R}}{1\over n+1} 
= {1\over n+1}{n+1\over n} 
= {1\over n}.
\label{firstsum}
\end{equation}
The first equality follows from the formula for an irreducible representation of $S_n$: $d_T={n!\over {\rm hooks}_T}$;
the second formula follows upon inserting the explicit expression for the dimensions of the hook Young diagrams $R$ and $R'$.
Thus, the coefficient of $f_{+R'}/f_{B}$ is $1/n$. Next consider
\begin{equation}
{\cal C}(n-1,N+M) = \sum_{(R',R'')} {{\rm hooks}_{R'}\over {\rm hooks}_{R''}}{1\over n^2}{f_{+R'}\over f_B}
= \sum_{(R',R'')} {d_{R''}\over d_{R'}}{1\over n}{f_{+R'}\over f_B} \, .
\label{secondsum}
\end{equation}
In the above sum, $R'$ runs over all hook Young diagrams with $n$ boxes and $R''$ runs over all possible ways of removing a box from 
this hook (there are 2 possible ways). The coefficient of $f_{+R'}/f_{B}$ in this last sum is
$$ \sum_{R''} {d_{R''}\over d_{R'}}{1\over n} = {1\over n}\, .$$
This follows because we sum over all possible subductions $R''$ of $R'$. This proves that (\ref{firstsum}) and (\ref{secondsum}) are
identical and hence we find the exact relation
\begin{equation}
{\cal C}_B(n,N)={\cal C} (n,N+M)- M {\cal C}(n-1,N+M)\, .
\label{amprel}
\end{equation}
We have explicitly checked that it holds for $n=1,2,3,4,5,6$. ${\cal C} (n,N+M)$ admits an expansions in ${1\over M+N}$; thus, we 
have demonstrated that the amplitudes ${\cal C}_B(n,N)$ admit an expansion in ${1\over M+N}$, albeit with the above extra $M$ dependence 
in the expansion. $M$ is the number of maximal giant gravitons making up the background; it is not surprising that there
are amplitudes that have some additional
dependence on $M$. The point is that all $N$ dependence in our amplitude has, according to (\ref{amprel}),
been replaced by an $N+M$ dependence which is perfectly consistent with what we found in the half-BPS sector.  

Given the above success one may ask if we can consider more general correlators and to work beyond the leading order. This 
more general analysis would tell us the class of operators for which our reorganization works, that is, the full class of operators that can 
be expanded in ${1\over M+N}$. Recall that the Schwinger-Dyson
equations of a theory determine the correlation functions of the theory. Thus, one possible approach to our problem (now that the above 
analysis has suggested what to search for), is to demonstrate the replacement $N\to M+N$ at the 
level of the Schwinger-Dyson equations. This demonstration turns out to be straightforward. Start in the trivial background; consider
first the simple Schwinger-Dyson equation\footnote{Since we compute correlators using (\ref{twopoint}), to obtain 
Schwinger-Dyson equations that determine our correlators we must consider a zero dimensional matrix model with action $S=\Tr (ZZ^\dagger)$.}
$$ 0=\int \left[dZdZ^\dagger\right] {d\over dZ_{ij}}\left( (Z^{n+1}Z^{\dagger\, n})_{ij}e^{-S}\right)\, .$$
Performing the derivative we find
$$ 
\left\langle \Tr (Z^{n+1} Z^{\dagger\, n+1})\right\rangle = N\left\langle \Tr (Z^{n} Z^{\dagger\, n})\right\rangle
+\sum_{r=1}^{n}\left\langle \Tr (Z^r)\Tr (Z^{n-r} Z^{\dagger\, n})\right\rangle\, .
$$
An easy computation (see Appendix A) shows that, in the background $B$, the above Schwinger-Dyson equation becomes
$$ 
\left\langle \Tr (Z^{n+1} Z^{\dagger\, n+1})\right\rangle_B = (N+M)\left\langle \Tr (Z^{n} Z^{\dagger\, n})\right\rangle_B
+\sum_{r=1}^{n}\left\langle \Tr (Z^r)\Tr (Z^{n-r} Z^{\dagger\, n})\right\rangle_B\, .
$$
Clearly, the net effect of the background is to replace $N\to N+M$ in perfect harmony with (\ref{amplituderelation}). This conclusion is rather
general: the net effect of the background, at the level of the Schwinger-Dyson equations,
is the replacement $N\to N+M$ suggesting that (\ref{amplituderelation}) should hold for all correlators of operators built
using only $Z$ and $Z^\dagger$. This conclusion is too quick. To determine the correlators, we should imagine solving these
equations iteratively; the value of $\left\langle \Tr (Z^{n+1} Z^{\dagger\, n+1})\right\rangle_B$ will thus
depend on $\left\langle \Tr (Z^{n} Z^{\dagger\, n})\right\rangle_B$. We should start the process with the equation that follows 
for $n=0$ 
$$ 
\left\langle \Tr (Z Z^{\dagger})\right\rangle_B = (N+M)N\, .
$$
The $N$ multiplying $(M+N)$ on the right hand side comes from $\Tr (1)$. This introduces an $N$ dependence, which is not replaced
by $M+N$. This result is in perfect agreement with (\ref{secondrelation}). By carefully analyzing the Schwinger-Dyson equations 
for the half-BPS loops, one can verify that a similar problem does not occur for these loops, which is consistent with (\ref{amplituderelation}).
The above departure from a pure $M+N$ dependence is rather mild and one may hope that there is a simple generalization of our results. 

We are claiming that we have a new ${1\over M+N}$ expansion parameter. Why not
simply replace $M+N\to (\mu +1)N$ with $\mu ={M\over N}$
so that we have the usual ${1\over N}$ expansion with $\mu$ dependent
coefficients? There are (at least) two reasons to reject this proposal

{\vskip 0.1cm}

\begin{itemize}

\item
The loop expansion parameter in the half-BPS sector is ${1\over M+N}$; the expansion coefficients have no additional $M$ or $N$
dependence. Since this sector includes gravitons, we should identify ${1\over M+N}$ as the $\hbar$ for the graviton interactions.
Indeed, a ${1\over N}$ expansion with $\mu$ dependent coefficients is an expansion whose coefficients change as $N$ is changed, 
indicating additional $\hbar$ dependence in these coefficients.

\item
Our relation (\ref{amplituderelation}) is exact and holds for any value of $M$. As the size of $M$ changes the character of the
expansion changes and it can be misleading to think that fluctuations are controlled by ${1\over N}$. If $M=O(1)$, $M+N$
can be replaced by $N$ and we have the usual ${1\over N}$ expansion as expected - this is the theory in the AdS$_5\times$S$^5$ background. 
In this case, $\mu =O({1\over N})$ and the coefficients again become $N$ independent. When $M=O(N)$, $M+N$ is itself of order $N$,
and ${1\over N}$ continues to control the size of fluctuations. This is the ${1\over M+N}$ expansion we found above - for example the theory 
in the LLM annulus background. In this case, $\mu =O(1)$, and the coefficients themselves are of order 1. When $M=O(N^2)$ we can replace
$M+N$ by $M$ so that our ${1\over M+N}$ expansion effectively becomes a ${1\over M}$ expansion. The correlators are dominated by 
contractions with the background and fluctuations are much smaller than in the usual ${1\over N}$ expansion: they are controlled by ${1\over M}\sim{1\over N^2}$. 
In this case, $\mu =O(N)$ so that the coefficients are unusually small - of size ${1\over N}$. Particularly in the last case, 
it would be absurd to suggest that we have a ${1\over N}$ expansion parameter.

\end{itemize}

{\vskip 0.1cm}

\noindent
This completes our demonstration that the BPS and near-BPS sectors of the theory admit a ${1\over M+N}$ expansion.

A study of the closed strings probing the LLM annulus background has been completed in \cite{Vazquez:2006id,Chen:2007gh,deMelloKoch,Koch:2008cm}.
One can describe the loops $O(\{n\})$ in terms of Cuntz oscillators (representing the $Z$s) populating a lattice (formed from the $Y$s) as in
\cite{Berenstein:2006qk,deMelloKoch:2007uu,deMelloKoch:2007uv,Bekker:2007ea}. The article \cite{Chen:2007gh} in particular, pointed out that the 
net effect of the background, on the one loop dilatation operator, was to rescale the Cuntz oscillators. This rescaling is nothing but the
relation we have found in (\ref{rescale})! This rescaling could also be accounted for by using a new 't Hooft coupling
$$ \lambda_{\rm new}=g_{YM}^2 (N+M). $$
This again looks natural: we have the inverse of the effective genus counting parameter times $g_{YM}^2$.
The net effect of the background on the one loop anomalous dimension operator (in this sector) is to replace $\lambda\to\lambda_{\rm new}$. At
present, we are investigating the effect of the background on the anomalous dimension operator at two loops\cite{busy}.

\section{Discussion}

One of the results of the papers \cite{Kimura:2007wy,Brown:2007xh,Bhattacharyya:2008rb}
is a new basis for the local gauge invariant operators of multimatrix models. This new basis
diagonalizes the two point functions and allows an exact computation of two point (and to some extent, multipoint) correlators
in the $g_{YM}^2=0$ limit. If in addition we consider near BPS correlators, corrections in $g_{YM}^2$ are suppressed, so that we 
obtain a rather complete description of these correlators. This allows us to ask and answer a number of interesting questions,
which may probe non-perturbative aspects of the dual quantum gravity.

In this article we have considered precisely such a question: the dynamics of operators whose classical dimension is $O(N^2)$.
These operators are dual to states that have a large mass and consequently back reaction on the dual geometry is important.
This is manifest in the fact that non-planar diagrams are no longer suppressed: although they come with the usual ${1\over N^2}$
suppressions, these are overpowered by huge combinatoric factors; the usual ${1\over N}$ expansion breaks down. In a number of
cases we have shown that it is possible to reorganize the expansion and have identified the small parameters that control these
expansions. These results were neatly captured by surprisingly simple relations between correlators in the trivial background and
correlators in the background of our heavy operator.

Another interesting result that we have obtained, is that in the multi ring backgrounds and in the multi matrix backgrounds
there was more than just one coupling. At first site this may seem puzzling, since the dilaton is a constant. To get some insight
into what is going on, consider QCD. At high energies QCD is well described by a Lagrangian of quarks and gluons
together with the coupling $g_{YM}^2$. At low energies, the coupling grows and one needs - somehow - to reorganize the theory 
of quarks and gluons into a low energy effective theory. The semi-classical objects in this low energy theory will be protons, neutrons, pions,...
and there will be many possible coupling constants telling us how these semi-classical objects interact. The relevance of this story
is that for the correlators we have studied, the planar approximation breaks down and one again needs to reorganize the theory. In this 
paper we have managed to explicitly perform this reorganization; the objects that we find that have different couplings are naturally 
interpreted as different semi-classical objects in the effective theory. 

There will be effects that are non-perturbative in the new expansion parameter. An example of a non-perturbative amplitude
is the transition from a maximal sphere giant graviton ($\chi_{[1^N]}(Z)$; $[1^N]$ denotes a Young diagram with one column
containing $N$ boxes) into KK gravitons with identical angular momentum $J << N$.
The amplitude is given by\cite{Corley:2001zk,Brown:2006zk} 
$$
  \frac{\left|\langle \chi_{[1^N]} (Z^{\dagger}) (\Tr (Z^J) )^{N/J}
    \rangle\right|^2}{\langle\chi_{[1^N]} (Z^\dagger )\chi_{[1^N]}
    (Z)\rangle\, \,\langle \Tr (Z^{\dagger J} )\Tr ( Z^J )
    \rangle^{N/J}} \sim (2\pi)^{\frac{1}{2}} e^{-{1\over g} - \frac{1}{2}
  \log(g) -(1/(gJ))\log(J) }
$$
where $g=1/N$. This is non-perturbative in $1/N$. To get the corresponding amplitude in the annulus, using Schur technology we can
again argue that we simply need to replace $N\to N+M$, so that the transition amplitude is non-perturbative in $1/(M+N)$.

The techniques of the article allow us to reorganize the $1/N$ expansion for classes of observables in backgrounds described by
Schur polynomials labeled by Young diagrams whose edges are all $O(N)$. This does not exhaust the operators that are dual to classical
geometries. Indeed, the Schur polynomials correspond to spacetimes whose LLM boundary condition preserve a rotational symmetry on the plane 
in which the LLM boundary condition is specified. This is a small subset of the possible 1/2 BPS LLM geometries. Further, even amongst those
LLM geometries with a rotationally invariant boundary condition, it is likely that we have only dealt with a subset of the geometries that 
can occur. Indeed, consider a triangular Young diagram $T$ such that its longest column has $N$ boxes, and each column has one box less than
the column to its left. Thus our Young diagram has $N$ columns in total. We have tested, in a number of cases, that factorization of the expectation
value
$$ 
\left\langle {\cal O}\right\rangle_{T}\equiv 
{\left\langle {\cal O}\chi_T(Z)\chi_T(Z^\dagger)\right\rangle\over\left\langle \chi_T(Z)\chi_T(Z^\dagger)\right\rangle}
$$
holds. This suggests that $\chi_T(Z)$ is again dual to a classical geometry. We have not yet understood how to reorganize the large $N$ expansion
in this case. 

One very interesting class of excitations of an operator with a very large classical dimension, is obtained by attaching ``open string
defects'' to the operator representing the background. Recent evidence\cite{Balasubramanian:2007bs,Fareghbal:2008ar}
suggests that these open string excited operators are dual to black hole micro states.
Consequently, we expect that black hole microstate dynamics is captured in the dynamics of these operators. Our results (see in
particular (\ref{rescale})) suggest that it may well be possible to write down spin chain descriptions for the open string excitations.
Towards this end, we have explored the proposed BMN-type sectors discovered in \cite{Fareghbal:2008ar}.
The existence of these limits is sure to play an important role in understanding
the gauge theory description of near-extremal charged black holes in AdS$_5$. Taking $M_1=O(\sqrt{N})$ and $M_2=O(\sqrt{N})$ with
the 't Hooft coupling $\lambda$ large but fixed, the charges $J_1$ and $J_2$ scale as $O(N^{3\over 2})$, whilst the difference 
$$ {\Delta -J_1-J_2 \over J_1+J_2}\sim N^{-{1\over 2}}\to 0 $$
in perfect agreement with the supergravity arguments of \cite{Fareghbal:2008ar}. Thus, strong and weak coupling results
agree - as expected for a BMN-like limit. 
We have further argued that if we hold $M_1=M_2=O(\sqrt{N})$ fixed we find an effective genus counting parameter $g_2=M$
and an effective 't Hooft coupling $\tilde{\lambda}= g_{YM}^2 M$. Keeping the effective 't Hooft coupling arbitrarily small,
the one loop correction to the anomalous dimension can be made arbitrarily small, even though $\lambda$ is sent to infinity
in the limit. This is similar to what happens in the BMN limit suggesting that 
it will be possible to compare perturbative field theory results to results obtained
from the dual gravitational description. The above effective 't Hooft coupling and genus counting parameters are naturally
identified with those of the gauge theory living on the world volume of our system of intersecting giant gravitons. It
would be very interesting to determine the dimensionality of this theory: 
is it a 2d gauge theory as the results of \cite{Balasubramanian:2007bs,Fareghbal:2008ar} imply?
Of course, the operators we have considered here are presumably too simple to describe the black hole microstates; for that one needs to
consider triangular Young diagrams \cite{Balasubramanian:2005mg}. The triangular Young diagrams have a large number of corners
which implies a large number of possible excitations of the operator. This is a significant increase in complexity as compared
to the studies in this paper. 

Finally, studying the rectangular Young diagrams that we have considered here will allow us to construct spin chains describing BMN loops
in the LLM annulus background. One can look for signatures of integrability for these spin chains. This provides a set up in which one can test
if integrability survives (a class of) non-planar corrections. We hope to report results in this direction in the near future\cite{busy}.

{\vskip 0.5cm}

\noindent
{\it Acknowledgements:} We would like to thank Shahin Sheikh-Jabbari for very helpful correspondence and Kevin Goldstein
for pleasant discussions. 
This work is based upon research supported by the South African Research Chairs Initiative 
of the Department of Science and Technology and National Research Foundation. Any opinion, findings and conclusions 
or recommendations expressed in this material are those of the authors and therefore the NRF and DST do not accept 
any liability with regard thereto. This work is also supported by NRF grant number Gun 2047219.

\appendix

\section{Schwinger-Dyson Equations in the Annulus Background}

The Schwinger-Dyson equations provide a powerful approach to computing correlators in the annulus background. They are
far more computationally efficient that the approach based on cutting rules developed in \cite{Koch:2008cm}. The advantage 
of the cutting rules are their generality: the cutting rules work in any background. In deriving the Schwinger-Dyson equation,
a crucial observation is that for the background we consider (recall that $B$ is a Young diagram with $M$ columns and $N$ rows)
$$\chi_B(Z)= \det (Z)^M\, .$$
We make repeated use of this fact and consequently, the results of this appendix apply only to the annulus background.

\subsection{Schwinger-Dyson Equations}

Start by considering
$$ 0=\int \left[dZdZ^\dagger\right] {d\over dZ_{ij}}\left( (Z^{n+1}Z^{\dagger\, n})_{ij}\chi_B(Z)\chi_B(Z^\dagger )e^{-S}\right)\, .$$
Carrying out the derivative is straightforward, except perhaps for the term obtained when the derivative acts on the background. To evaluate 
this term, note that
$$ {d\over dZ_{ij}}\chi_B(Z)= {d\over dZ_{ij}}\det (Z)^M =M(Z^{-1})_{ji}\det (Z)^M\, .$$
Thus, this term contributes
$$M\left\langle \Tr (Z^n Z^{\dagger\, n})\right\rangle_B$$
to the Schwinger-Dyson equation. Next, focus on the term obtained by acting on the first $Z$ in $(Z^{n+1}Z^{\dagger\, n})_{ij}$ which gives
$$N\left\langle \Tr (Z^n Z^{\dagger\, n})\right\rangle_B\, ;$$
these two terms combine to give the claimed $N\to M+N$ replacement in the Schwinger-Dyson equations. Writing out all of the terms we have
$$
\left\langle \Tr (Z^{n+1} Z^{\dagger\, n+1})\right\rangle_B = (N+M)\left\langle \Tr (Z^{n} Z^{\dagger\, n})\right\rangle_B
+\sum_{r=1}^{n}\left\langle \Tr (Z^r)\Tr (Z^{n-r} Z^{\dagger\, n})\right\rangle_B\, .
$$
A slightly more general Schwinger-Dyson equation which we found useful in the computation of correlators reads
$$ 0=\int \left[dZdZ^\dagger\right] {d\over dZ_{ij}}\left( (Z^m Z^{\dagger\, n})_{ij}\Tr (Z^{n+1-m})\chi_B(Z)\chi_B(Z^\dagger )e^{-S}\right)\, $$
which is easily seen to give
$$
\left\langle \Tr (Z^{\dagger\, n+1}Z^m)\Tr (Z^{n+1-m})\right\rangle_B=\sum_{r=1}^{m-1}
\left\langle \Tr (Z^{\dagger\, n}Z^{m-r-1})\Tr (Z^r)\Tr (Z^{n+1-m})\right\rangle_B$$
$$+(N+M)\left\langle \Tr (Z^{\dagger\, n}Z^{m-1})\Tr (Z^{n+1-m})\right\rangle_B
+(n+1-m)\left\langle \Tr (Z^{n} Z^{\dagger\, n})\right\rangle_B\, .
$$
This starting point could easily be generalized to
$$ 0=\int \left[dZdZ^\dagger\right] {d\over dZ_{ij}}\left( (Z^m Z^{\dagger\, n})_{ij}\Tr (Z^{n+1-m})O(Z,Z^\dagger )\chi_B(Z)\chi_B(Z^\dagger )e^{-S}\right)\, $$
where $O(Z,Z^\dagger )$ is any gauge invariant operator.

Computing correlators is now straightforward. We can obtain all correlators of the form 
$\left\langle \prod_{i,j}\Tr (Z^{n_i})\Tr ((Z^\dagger)^{m_j})\right\rangle_B$ using 
(\ref{amplituderelation}). Using these in the above Schwinger-Dyson equations, we can easily
determine the correlators $\left\langle \prod_i \Tr (Z^{n_i}Z^{\dagger \, n_i})\right\rangle_B$
which are of relevance for the near-BPS sector of the theory.

\subsection{Testing Factorization}

Now that we can use the Schwinger-Dyson equations to compute correlators exactly, we can answer some interesting
questions. One obvious question is if the annulus geometry provides a good background. For this to be the case,
we need to have a factorization of the background expectation values of gauge invariant observables. This implies
that a single saddle point is dominating the gauge theory path integral. By the gauge theory/gravity correspondence, 
this saddle point represents a particular space-time geometry in gravity, that is, a classical spacetime has emerged.
One nice general result follows from
$$ 0=\int \left[dZdZ^\dagger\right] {d\over dZ_{ij}}\left( Z_{ij}\big[\Tr (ZZ^\dagger )\big]^n\chi_B(Z)\chi_B(Z^\dagger )e^{-S}\right)\, $$
which implies that
$$ \left\langle \big[\Tr (ZZ^\dagger )\big]^{n+1}\right\rangle_B = (N^2+MN+n)\left\langle \big[\Tr (ZZ^\dagger )\big]^{n}\right\rangle_B\, .$$
This recursion relation is easily solved to give
$$ \left\langle \big[\Tr (ZZ^\dagger )\big]^{n+1}\right\rangle_B = \prod_{i=0}^n (N^2+MN+i)\, .$$
Keeping only the leading order\footnote{Of course, $n$ and $p$ (used below) are $O(1)$.}, we have
$$ \left\langle \big[\Tr (ZZ^\dagger )\big]^{n+1}\right\rangle_B = (N^2+MN)^{n+1} =\left\langle \Tr (ZZ^\dagger )\right\rangle_B^{n+1}\, ,$$
demonstrating factorization for these amplitudes. We can easily generalize this by considering
$$ 0=\int \left[dZdZ^\dagger\right] {d\over dZ_{ij}}\left( Z_{ij}\Tr (Z^p Z^{\dagger p})\big[\Tr (ZZ^\dagger )\big]^n\chi_B(Z)\chi_B(Z^\dagger )e^{-S}\right)\, $$
which implies
$$\left\langle \big[\Tr (ZZ^\dagger )\big]^{n+1}\Tr (Z^p Z^{\dagger p})\right\rangle_B =
(N^2+MN+n+p)\left\langle \big[\Tr (ZZ^\dagger )\big]^{n}\Tr (Z^p Z^{\dagger p})\right\rangle_B\, .$$
Once again this is easy to solve, giving
$$ \left\langle \big[\Tr (ZZ^\dagger )\big]^{n+1}\Tr (Z^p Z^{\dagger p})\right\rangle_B =
\prod_{i=0}^n (N^2+MN+i+p)\left\langle \Tr (Z^p Z^{\dagger p})\right\rangle_B\, ,$$
which becomes, at the leading order,
\begin{eqnarray} 
\left\langle \big[\Tr (ZZ^\dagger )\big]^{n+1}\Tr (Z^p Z^{\dagger p})\right\rangle_B &=&
(N^2+MN)^{n+1}\left\langle \Tr (Z^p Z^{\dagger p})\right\rangle_B\nonumber\\
&=& \left\langle \Tr (Z^p Z^{\dagger p})\right\rangle_B
\left\langle \Tr (ZZ^\dagger )\right\rangle_B^{n+1}\, ,\nonumber
\end{eqnarray}
again demonstrating factorization.

Above we have been careful to compute things to all orders. If we simply assume factorization and keep only the leading order,
we get additional information about the leading behavior of various loops. For example, 
$$\left\langle \Tr (Z^{p+1} Z^{\dagger\, p+1})\prod_{i}\Tr (Z^{n_i}Z^{\dagger\, n_i})\right\rangle_B =
(N+M)\left\langle \Tr (Z^p Z^{\dagger\, p})\prod_{i}\Tr (Z^{n_i}Z^{\dagger\, n_i})\right\rangle_B
$$
$$ 
+\sum_{r=1}^{p}\left\langle \Tr (Z^{p-r} Z^{\dagger\, p})\Tr (Z^r)\prod_{i}\Tr (Z^{n_i}Z^{\dagger\, n_i}) \right\rangle_B
$$
$$
+\sum_j \sum_{r=0}^{n_j-1}
\left\langle \Tr(Z^{r+p+1} Z^{\dagger\, p} Z^{n_j-r-1}Z^{\dagger\, n_j})\prod_{i\ne j}\Tr (Z^{n_i}Z^{\dagger\, n_i})\right\rangle_B ,$$
becomes, after assuming factorization
$$\left\langle \Tr (Z^{p+1} Z^{\dagger\, p+1})\right\rangle_B\prod_{i}\left\langle\Tr (Z^{n_i}Z^{\dagger\, n_i})\right\rangle_B =
(N+M)\left\langle \Tr (Z^p Z^{\dagger\, p})\right\rangle_B\prod_{i}\left\langle\Tr (Z^{n_i}Z^{\dagger\, n_i})\right\rangle_B
$$
$$
+\sum_j \sum_{r=0}^{n_j-1}
\left\langle \Tr(Z^{r+p+1} Z^{\dagger\, p} Z^{n_j-r-1}Z^{\dagger\, n_j})\right\rangle_B
\prod_{i\ne j}\left\langle\Tr (Z^{n_i}Z^{\dagger\, n_i})\right\rangle_B ,$$
The second term of the right hand side is subleading compared to the first term because (i) the first term is multiplied by $(N+M)$ and
(ii) the second term has one less trace in it. Thus it may be dropped to give
$$\left\langle \Tr (Z^{p+1} Z^{\dagger\, p+1})\right\rangle_B\prod_{i}\left\langle\Tr (Z^{n_i}Z^{\dagger\, n_i})\right\rangle_B =
(N+M)\left\langle \Tr (Z^p Z^{\dagger\, p})\right\rangle_B\prod_{i}\left\langle\Tr (Z^{n_i}Z^{\dagger\, n_i})\right\rangle_B\, .
$$
Iterating this relation, the above result is clearly equivalent to
$$ 
\left\langle \Tr (Z^q Z^{\dagger\, q})\right\rangle_B = N(N+M)^q \, .
$$
which we know is correct.

\subsection{Testing the Cutting Rules}

In \cite{Koch:2008cm} a method to compute general correlators in any arbitrary LLM background was given. 
Now that we have an efficient way to compute correlators in the annulus background we can ask: Do the cutting rule methods of
\cite{Koch:2008cm} really work? In this appendix we will compute a specific correlator, first using the Schwinger-Dyson equations
and then using the cutting rules. We have found complete agreement between the cutting rule result and the result from the
Schwinger-Dyson equations for any correlator we have computed. The cutting rules work.

Starting from
$$ 0=\int \left[dZdZ^\dagger\right] {d\over dZ_{ij}}\left( (Z^{\dagger\, n}Z^m Z^{\dagger\, p}Z^{n+p-m+1})_{ij}\chi_B(Z)\chi_B(Z^\dagger )e^{-S}\right)\, $$
we obtain
$$ \left\langle \Tr (Z^{\dagger n+1}Z^m Z^{\dagger p}Z^{n+p-m+1})\right\rangle_B =
\sum_{r=0}^{m-1}\left\langle \Tr (Z^{\dagger n}Z^r)\Tr( Z^{\dagger p}Z^{n+p-r})\right\rangle_B $$
$$+\sum_{r=0}^{n+p-m-1}\left\langle \Tr (Z^{\dagger n}Z^m Z^{\dagger p}Z^r)\Tr (Z^{n+p-m-r})\right\rangle_B
+(N+M)\left\langle \Tr (Z^{\dagger n}Z^m Z^{\dagger p}Z^{n+p-m})\right\rangle_B \, .$$
Setting $n=0$, $m=1$ and $p=1$ we have
$$\left\langle \Tr (Z^{\dagger}ZZ^{\dagger}Z)\right\rangle_B=(2N+M)\left\langle \Tr (Z^{\dagger }Z)\right\rangle_B
=(2N+M)N(N+M)\, .$$

We will summarize the cutting rule computation; for more details the reader should consult \cite{Koch:2008cm}.
Evaluating this correlator using cutting rules, there are four contributions: the term with no contractions
with the background gives $2N^3$. The terms coming from contracting one $Z$ in the loop with a $Z^\dagger$ in
$\chi_B(Z^\dagger)$ give
$$ 4N \left\langle \Tr \left({d\over dZ}{d\over dZ^\dagger}\right)\chi_B(Z)\chi_B(Z^\dagger)\right\rangle =4N^2 M . $$
The terms coming from contracting both $Z$s in the loop with $Z^\dagger$s in $\chi_B(Z^\dagger)$ give
$$ \left\langle \Tr \left({d\over dZ}{d\over dZ^\dagger}{d\over dZ}{d\over dZ^\dagger}\right)\chi_B(Z)\chi_B(Z^\dagger)\right\rangle 
=M^2N- N^2 M . $$
To evaluate this last contribution we had to cut the trace of four derivatives into a product of two traces, each containing two derivatives.
This is accompanied by a nontrivial trace insertion factor that we evaluated in representation $B$. Summing these terms we have
$$ 2N^3+4N^2 M+M^2N-N^2 M = (2N+M)N(N+M)\, .$$

\section{Schwinger-Dyson Equations for $>1$ Charge Background}

The background of interest in this appendix is (recall that $r_1$ is a rectangular Young diagram with $N$ rows and $M_1$ columns
and $r_2$ is a rectangular Young diagram with $N$ rows and $M_2$ columns)
$$ \chi_{r_1}(Z)\chi_{r_2}(Y)= (\det (Z))^{M_1}(\det (Y))^{M_2}\, .$$
The Schwinger-Dyson equations continue to provide a powerful approach to correlator computations, when we consider this background built
using more than one matrix. In this appendix we will give a few example computations. Consider the
identity\footnote{The action $S=\Tr (ZZ^\dagger)+\Tr (YY^\dagger)$.}
$$0=\int \big[ dZdZ^\dagger dYdY^\dagger\big] {d\over dZ_{ij}}\left( (Z^n Y^m Y^{\dagger\, m}Z^{\dagger\, n-1})_{ij}
\chi_{r_1}(Z)\chi_{r_1}(Z^\dagger)\chi_{r_2}(Y)\chi_{r_2}(Y^\dagger)e^{-S}\right)$$
which leads to the following Schwinger-Dyson equation
$$ 
\left\langle \Tr (Z^n Y^m Y^{\dagger\, m}Z^{\dagger\, n})\right\rangle_{(r_1,r_2)}= 
(N+M_1)\left\langle \Tr (Z^{n-1} Y^m Y^{\dagger\, m}Z^{\dagger\, n-1})\right\rangle_{(r_1,r_2)}
$$
$$
+\sum_{r=1}^{n-1}\left\langle \Tr (Z^r)\Tr (Z^{n-1-r} Y^m Y^{\dagger\, m}Z^{\dagger\, n-1})\right\rangle_{(r_1,r_2)}\, .
$$
If we had been working in the trivial vacuum, the only difference would have been to replace $N+M_1$ in the above equation by $N$.
One way to think about the above Schwinger-Dyson equation is that to go from the left hand side to the right hand side, we perform 
one of the $Z$ Wick contractions. To obtain the equation that follows when we perform a $Y$ Wick contraction, start with the identity
$$0=\int \big[ dZdZ^\dagger dYdY^\dagger\big] {d\over dY_{ij}}\left( (Y^n Z^m Z^{\dagger\, m}Y^{\dagger\, n-1})_{ij}
\chi_{r_1}(Z)\chi_{r_1}(Z^\dagger)\chi_{r_2}(Y)\chi_{r_2}(Y^\dagger)e^{-S}\right)$$
which leads to the following Schwinger-Dyson equation
$$ 
\left\langle \Tr (Y^n Z^m Z^{\dagger\, m}Y^{\dagger\, n})\right\rangle_{(r_1,r_2)}= 
(N+M_2)\left\langle \Tr (Y^{n-1} Z^m Z^{\dagger\, m}Y^{\dagger\, n-1})\right\rangle_{(r_1,r_2)}
$$
$$
+\sum_{r=1}^{n-1}\left\langle \Tr (Y^r)\Tr (Y^{n-1-r} Z^m Z^{\dagger\, m}Y^{\dagger\, n-1})\right\rangle_{(r_1,r_2)}\, .
$$
If we had been working in the trivial vacuum, the only difference would have been to replace $N+M_2$ in the above equation by $N$.
This structure is parallel to the structure we found for backgrounds constructed using a single matrix: in this case we have found that
to reproduce correlators of operators built only using $Z$s or $Z^\dagger$s ($Y$s or $Y^\dagger$s) we simply replace $N\to N+M_1$
($N\to N+M_2$). This structure is again emerging at the level of the Schwinger-Dyson equations. 

Consider the last Schwinger-Dyson equation given above. In the large $N$ limit the first term on the right hand side gives the leading 
contribution. The second term has one more trace in it whilst the first term is multiplied by $(N+M_2)$. Naive counting of powers of $N$
would suggest that these two terms are of the same order. However, the leading contribution to the second term 
$\left\langle \Tr (Y^r)\right\rangle_{(r_1,r_2)}\left\langle\Tr (Y^{n-1-r} Z^m Z^{\dagger\, m}Y^{\dagger\, n-1})\right\rangle_{(r_1,r_2)}$
vanishes. Dropping the second term and iterating we find
\begin{equation} 
\left\langle \Tr (Y^n Z^m Z^{\dagger\, m}Y^{\dagger\, n})\right\rangle_{(r_1,r_2)}= 
(N+M_2)^n\left\langle \Tr (Z^m Z^{\dagger\, m})\right\rangle_{(r_1,r_2)}=
N(N+M_1)^m (N+M_2)^n\, .
\label{sd}
\end{equation}
This looks very similar to the relation (\ref{rescale}); indeed, we can write
$$
\left\langle O(Z,Z^\dagger,Y,Y^\dagger)\right\rangle_{(r_1,r_2)} 
=\left\langle O\left(\sqrt{N+M_1\over N}Z,\sqrt{N+M_1\over N}Z^\dagger,
\sqrt{N+M_2\over N}Y,\sqrt{N+M_2\over N}Y^\dagger\right)\right\rangle\, .
$$
Relations of this type would again be very useful in deriving spin chains for loops in this two matrix background.

Finally, although we have already argued that there are large 't Hooft coupling corrections to the background, it is still interesting
to ask if factorization holds. Such backgrounds are naturally interpreted as classical backgrounds that receive curvature corrections.
For operators that do not mix $Z$s and $Y$s in the same trace, correlators factorize into a $Z$ correlator times a $Y$ correlator. Using
the results of Appendix A we clearly have factorization in this case. For operators with mixed traces, a bit more work is needed.
We will give some examples in which factorization is clear. Consider
$$
{\small
0=\int \big[ dZdZ^\dagger dYdY^\dagger\big] {d\over dY_{ij}}\left( (Y^{n+1} Z^m Z^{\dagger\, m}Y^{\dagger\, n})_{ij}\times \right.}
$$
$${\small \left. \times \prod_a \Tr (Z^{n_a}Z^{\dagger n_a})
\chi_{r_1}(Z)\chi_{r_1}(Z^\dagger)\chi_{r_2}(Y)\chi_{r_2}(Y^\dagger)e^{-S}\right)}
$$
which implies
$$
\left\langle \Tr (Y^{\dagger\, n} Y^{n}Z^m Z^{\dagger m})\prod_a \Tr (Z^{n_a}Z^{\dagger n_a})\right\rangle_{(r_1,r_2)}
$$
$$
= (N+M_2)\left\langle \Tr (Y^{\dagger\, n-1} Y^{n-1}Z^m Z^{\dagger m})\prod_a \Tr (Z^{n_a}Z^{\dagger n_a})\right\rangle_{(r_1,r_2)}
$$
$$
+\sum_{r=1}^{n-1}\left\langle \Tr (Y^r)\Tr (Y^{n-1-r} Z^m Z^{\dagger\, m}Y^{\dagger\, n-1})\prod_a \Tr (Z^{n_a}Z^{\dagger n_a})
\right\rangle_{(r_1,r_2)}\, .
$$
In the large $N$ limit the first term on the right hand side gives the leading contribution so that we can drop the second term.
Iterating, we find
$$
\left\langle \Tr (Y^{\dagger\, n} Y^{n}Z^m Z^{\dagger m})\prod_a \Tr (Z^{n_a}Z^{\dagger n_a})\right\rangle_{(r_1,r_2)}
%
= (N+M_2)^n \left\langle \Tr (Z^m Z^{\dagger m})\prod_a \Tr (Z^{n_a}Z^{\dagger n_a})\right\rangle_{(r_1,r_2)}
$$
$$
= (N+M_2)^n\left\langle \Tr (Z^m Z^{\dagger m})\right\rangle_{(r_1,r_2)} 
\prod_a \left\langle\Tr (Z^{n_a}Z^{\dagger n_a})\right\rangle_{(r_1,r_2)}\, ,
$$
where to get the last equality we used factorization of the $Z,Z^\dagger$ correlators. Now, use (\ref{sd}) to identify
$(N+M_2)^n \left\langle \Tr (Z^m Z^{\dagger m})\right\rangle$ as $\left\langle \Tr (Y^{\dagger n}Y^n Z^m Z^{\dagger m})\right\rangle$
in the last line above so that
$$
\left\langle \Tr (Y^{\dagger n} Y^n Z^m Z^{\dagger m})\prod_a \Tr (Z^{n_a}Z^{\dagger n_a})\right\rangle_{(r_1,r_2)}
$$
$$
= \left\langle \Tr (Y^{\dagger n} Y^n Z^m Z^{\dagger m})\right\rangle_{(r_1,r_2)} 
\prod_a \left\langle \Tr (Z^{n_a}Z^{\dagger n_a})\right\rangle_{(r_1,r_2)}\, .
$$

We can give a rather general argument for factorization: consider
$$ 
{\small
0=\int \big[ dZdZ^\dagger dYdY^\dagger\big] {d\over dY_{ij}}\left( (Y^{p+1} Z^n Z^{\dagger\, n}Y^{\dagger\, p})_{ij}{\cal O}
\chi_{r_1}(Z)\chi_{r_1}(Z^\dagger)\chi_{r_2}(Y)\chi_{r_2}(Y^\dagger)e^{-S}\right)}
$$
where ${\cal O}$ is any gauge invariant operator. This implies
$$
\left\langle \Tr (Y^{\dagger\, p+1} Y^{p+1}Z^n Z^{\dagger n}){\cal O}\right\rangle_{(r_1,r_2)}
= (N+M_2)\left\langle \Tr (Y^{\dagger\, p} Y^{p}Z^n Z^{\dagger n}){\cal O}\right\rangle_{(r_1,r_2)}
$$
$$
+\sum_{r=1}^p\left\langle \Tr (Y^r)\Tr (Y^{p-r} Z^m Z^{\dagger\, m}Y^{\dagger\, p}){\cal O}
\right\rangle_{(r_1,r_2)}
+
\left\langle  (Y^{p+1} Z^n Z^{\dagger\, n}Y^{\dagger\, p})_{ij}{d\over dY_{ij}}{\cal O}
\right\rangle_{(r_1,r_2)}\, .
$$
The second term on the right hand side can be dropped - it vanishes at leading order. The third term on the right hand side can also be dropped - it
represents a loop joining term, so that it has one less trace than the first term. If we now rescale $Y\to {N\over N+M_2}Y$ and
$Z\to {N\over N+M_1}Z$ we recover the Schwinger-Dyson equations of the theory in the trivial vacuum (after dropping the same two terms justified with 
the same two reasons). We know that factorization was a property of the old Schwinger-Dyson equations so that we have just learnt that it is a property 
of the new Schwinger-Dyson equations too.

\section{Anomalous Dimension for $>1$ Charge Background}

In this section we will explain how to compute the one loop anomalous dimension of the backgrounds considered in section 2.4. 

\subsection{An Identity: Excited Giant Correlators}

Correlation functions of restricted Schur polynomials have been computed in \cite{deMelloKoch:2007uu} (see also \cite{deMelloKoch:2007uv,Bekker:2007ea}).
The logic in these computations is first to contract the open string words and then to compute the remaining contractions. In this section we will obtain
a formula that describes the result of contracting all fields {\it except} the open string words. Although we derive our formula for the case of one
string attached, it is simple to extend it to the general case. The formula we are after says
$$ \left\langle \chi_{R,R'}^{(1)}(Z,W)\chi_{R,R'}^{(1)}(Z^\dagger ,W^\dagger)\right\rangle = A \left\langle \Tr (WW^\dagger)\right\rangle
                                                                                           + B \left\langle \Tr (W)\Tr (W^\dagger)\right\rangle\, .$$
Recall\cite{deMelloKoch:2007uu} that the allowed index structure for open string word two point functions is
$$ \left\langle W^i_j (W^\dagger)^k_l\right\rangle =\delta^i_l\delta^k_j F_0 +\delta^i_j\delta^k_l F_1 \, .$$
Thus,
$$ \left\langle \chi_{R,R'}^{(1)}(Z,W)\chi_{R,R'}^{(1)}(Z^\dagger ,W^\dagger)\right\rangle = A (N^2 F_0 + N F_1)
                                                                                           + B (N^2 F_1 + N F_0)\, .$$
From the technology developed in \cite{deMelloKoch:2007uu}, we also know that
$$ \left\langle \chi_{R,R'}^{(1)}(Z,W)\chi_{R,R'}^{(1)}(Z^\dagger ,W^\dagger)\right\rangle = 
      {{\rm hooks}_R\over {\rm hooks}_{R'}} f_R F_0 + c_{RR'} f_R F_1 \, . $$
It is now trivial to find
$$ A=\left({{\rm hooks}_R\over {\rm hooks}_{R'}}N^2 -c_{RR'}N\right){f_R\over N^4 -N^2}\, ,$$
$$ B=\left(N^2 c_{RR'}-N{{\rm hooks}_R\over {\rm hooks}_{R'}}\right){f_R\over N^4 -N^2}\, .$$

\subsection{Leading Contribution to Background Correlator}

The formulas we write in this section will not be general; we are considering the backgrounds $r_1$ and $r_2$ of section 2.4. With a little extra effort
one could be general. We are interested in computing the normalized correlation function
$$ \left\langle \chi_{r_1}(Z^\dagger)\chi_{r_2}(Y^\dagger)\chi_{r_1}(Z)\chi_{r_2}(Y)\right\rangle\, , $$
to one loop. According to Appendix B of \cite{Constable:2002hw}, the D-term, self energy and gluon exchange cancel at one loop order (using techniques of 
\cite{D'Hoker:1998tz}), so to this order we only need to consider the contributions from the F-term. Towards this end we will now evaluate
$$I_1= \left\langle \chi_{r_1}(Z^\dagger)\chi_{r_2}(Y^\dagger)\chi_{r_1}(Z)\chi_{r_2}(Y)\Tr (\big[Z,Y\big]\big[Z^\dagger,Y^\dagger])\right\rangle\, . $$
The tricky part of this computation is the evaluation of the color combinatoric factor. To do this evaluation we can work in zero dimensions.
Since we drop self energy corrections, the F-term is normal ordered and hence the above correlator can be written as
$$ \left\langle \Tr (\big[{\partial\over \partial Z},{\partial\over\partial Y}\big]
\big[{\partial\over\partial Z^\dagger},{\partial\over\partial Y^\dagger}])
\chi_{r_1}(Z^\dagger)\chi_{r_2}(Y^\dagger)\chi_{r_1}(Z)\chi_{r_2}(Y)\right\rangle\, . 
$$
This can be rewritten, using dummy open string variables, as
$$ \left\langle \Tr (\big[{\partial\over \partial W},{\partial\over\partial V}\big]
\big[{\partial\over\partial W^\dagger},{\partial\over\partial V^\dagger}])
\chi_{r_1,r_1'}^{(1)}(Z^\dagger,W^\dagger)\chi_{r_2,r_2'}^{(1)}(Y^\dagger,V^\dagger)
\chi^{(1)}_{r_1,r_1'}(Z,W)\chi^{(1)}_{r_2,r_2'}(Y,V)\right\rangle\, . 
$$
For both $r_1$ and $r_2$ there is only one way possible to attach the open string. Using the results of the previous subsection we now obtain
$$\Tr (\big[{\partial\over \partial W},{\partial\over\partial V}\big]
\big[{\partial\over\partial W^\dagger},{\partial\over\partial V^\dagger}])
\left( A_1 \Tr (WW^\dagger) + B_1 \Tr (W)\Tr (W^\dagger)\right)
\left( A_2 \Tr (VV^\dagger) + B_2 \Tr (V)\Tr (V^\dagger)\right),$$
where
$$ A_1={M_1\over N}f_{r_1},\quad A_2={M_2 \over N}f_{r_2},\quad B_i=0\, .$$
It is a simple matter to find
$$ {I_1\over f_{r_1}f_{r_2}} = -2NM_1 M_2+2{M_1 M_2\over N}\, .$$
The leading contribution to this correlator comes from the terms
$$ \Tr (ZYY^\dagger Z^\dagger)+\Tr (YZZ^\dagger Y^\dagger)\, .$$
This computation will allow us, in the next subsection, to identify and evaluate the leading contribution to the one loop anomalous dimension.

\subsection{Leading Contribution to the One Loop Dilatation Operator}

The $O(g_{YM}^0)$ contribution to the anomalous dimension
$$ D_0 = \Tr \left( Z{\partial\over\partial Z}\right) + \Tr \left( Y{\partial\over\partial Y}\right)$$
gives $\Delta_0 = N M_1+N M_2$. To obtain the leading piece of the $O(g_{YM}^2)$ contribution to the anomalous dimension, which
we have identified in the previous subsection, replace
$$ D_1 =-2g_{YM}^2\Tr :\Big[ {\partial\over\partial Y},{\partial\over\partial Z}\Big]\Big[ Y,Z\Big]:\to 
2g_{YM}^2\Tr\left(ZY{\partial\over\partial Y}{\partial\over\partial Z}\right)+
2g_{YM}^2\Tr\left(YZ{\partial\over\partial Z}{\partial\over\partial Y}\right)\, .$$
The normal ordering symbols here indicate that derivatives within the normal ordering symbols do not act on fields inside the normal
ordering symbols.
It is now straightforward to argue that
$$ 
\left(2g_{YM}^2\Tr\left(ZY{\partial\over\partial Y}{\partial\over\partial Z}\right)+
2g_{YM}^2\Tr\left(YZ{\partial\over\partial Z}{\partial\over\partial Y}\right)\right)\chi_{r_1}(Z)\chi_{r_2}(Y)$$
$$=4g_{YM}^2 NM_1 M_2 \chi_{r_1}(Z)\chi_{r_2}(Y) \, .$$
We are interested in two cases:

\begin{itemize}

\item{} Both $M_1$ and $M_2$ are $O(N)$. Holding $g_{YM}^2 N=\lambda$ fixed and large (which is the regime in which we expect that 
          we can trust the dual geometry), we find a one loop correction of $O(N^2)$ times $\lambda$ to the tree level value which is
          itself $O(N^2)$. At large $N$ the tree level and one loop results are of the same order.

\item{} Both $M_1$ and $M_2$ are $O(\sqrt{N})$. Holding $g_{YM}^2 N=\lambda$ fixed and large, we find a one loop correction of $O(N)$
          times $\lambda$ to the tree level value which is itself $O(N^{3/2})$. At large $N$ the tree level result dominates the one
          loop correction.

\end{itemize}

\noindent
Notice that if we take
$M_1$ to be $O(N)$ and keep $M_2$ to be $O(1)$ or if we take $M_2$ to be $O(N)$ and keep $M_1$ to be $O(1)$, the one loop correction 
becomes negligible as compared to the tree level value, as we would expect. It is also interesting to note that our operator is
an eigenoperator of $D_1$, so that at one loop and at large $N$ it does not mix with other operators.

\section{Notation and Useful Results}

If $R$ is a Young diagram, $R'$ is a Young diagram obtained by removing one box from $R$.

We have used the weight\footnote{These are not the Dynkin weights. We use the terms ``weight of a box'' and ``factor of a box''
interchangeably.} and the hook of each box in a Young diagram. A box in row $i$ and column 
$j$ has a weight equal to $N-i+j$. We use the notation $f_R$ to denote the product of the weights of Young diagram $R$. For example
$$ f_{\tiny \yng(2,1)}=N(N+1)(N-1)\, .$$
We use $c_{RR'}$ to denote the weight of the box that must be dropped from $R$ to obtain $R'$.

To obtain the hook associated to a given box, draw a line starting from the given box towards the bottom of
the page until you exit the Young diagram, and another line starting from the same box towards the right until
you again exit the diagram. These two lines form an elbow - what we call the hook. The hook length for the given 
box is obtained by counting the number of boxes the elbow belonging to the box passes through. The notation
${\rm hooks}_R$ means the product of hook lengths of Young diagram $R$. Thus, for example
$${\rm hooks}_{\tiny \yng(2,1)}=1\cdot 1\cdot 3\, .$$

We have made extensive use of the exact two point function of the Schur polynomial\cite{Corley:2001zk}
$$ \left\langle\chi_R(Z^\dagger)\chi_S(Z)\right\rangle = \delta_{RS}f_R $$
where $\delta_{RS}$ is one if $R$ and $S$ are identical Young diagrams and zero otherwise. 
We have also used the product rule
$$\chi_R(Z)\chi_S(Z)=\sum_T g_{RST}\chi_T(Z),$$
where $g_{RST}$ is the Littlewood-Richardson number.

\end{document}